\documentclass[preprint, superscriptaddress, amsmath,amssymb]{revtex4-2}
\usepackage{verbatim} 
\usepackage{graphicx}
\usepackage{dcolumn}
\usepackage{bm}
\usepackage{amsmath, amssymb}
\usepackage{array} 
\usepackage{algorithm}
\usepackage{algpseudocode} 
\usepackage{enumitem}
\usepackage{xcolor}
\usepackage{placeins}
\usepackage{tabularx}
\usepackage{algorithm}
\usepackage{algpseudocode}
\usepackage{amsmath}
\usepackage{hyperref}

\usepackage{framed}
\newtheorem{alga}{Algorithm}{\scshape}{\sffamily}
\newenvironment{algo}
  {\begin{framed}\begin{alga}\begin{em}\begin{sffamily}}
  {\end{sffamily}\end{em}\end{alga}\end{framed}}
\usepackage{setspace}

\renewcommand{\vss}{\vspace{0.25cm}}

\begin{document}

\title{Parallel adaptive reweighting importance sampling for Bayesian astrophysics}

\author{Miaoxin Liu}
\affiliation{Department of Physics, National University of Singapore, Singapore 117551}
\author{Alvin J. K. Chua}%
\email{alvincjk@nus.edu.sg}
\affiliation{Department of Physics, National University of Singapore, Singapore 117551}
\affiliation{Department of Mathematics, National University of Singapore, Singapore 119076}


\date{\today}

\begin{abstract}
Efficient sampling from high-dimensional, multi-modal posteriors is a central challenge in Bayesian inference across the physical sciences. Gravitational-wave (GW) astronomy, where likelihood evaluations are computationally expensive and posteriors are complex, is one application motivating this work. Popular families of methods like Markov-chain Monte Carlo, nested sampling, and importance sampling all rely on proposal distributions to guide exploration. Because prior knowledge of the target is often limited, practitioners can adopt adaptive proposals that iteratively refine themselves using information gained from previously drawn samples. Traditional adaptive strategies, however, struggle in high-dimensional multi-modal settings: complex, non-linear correlations are hard to capture, and hyperparameters typically require tedious, problem-specific tuning.
To address these issues, we introduce Parallel Adaptive Reweighting Importance Sampling (PARIS; descriptively, ``seed sampling''). PARIS models its proposal as a Gaussian mixture whose component centers are the existing samples and whose component weights match the current importance weights. New draws from the proposal therefore concentrate around high-weight regions, while candidate points in unexplored areas receive intentionally inflated weights. As the algorithm continuously reweights all samples up to the latest proposal, any initial over-weighting self-corrects once additional neighbor samples are collected.
To enable rapid reweighting, we present an efficient update scheme and evaluate PARIS on illustrative toy problems and more realistic gravitational-wave parameter estimation tasks. PARIS achieves accurate posterior reconstruction and evidence estimation with substantially fewer function evaluations than competing approaches, highlighting its promise for widespread use in astrophysical data analysis.
\end{abstract}



\section*{Keywords}
gravitational waves, Bayesian inference, importance sampling

\maketitle

\section{Introduction}

Bayesian inference is the standard framework for parameter estimation, uncertainty quantification and model comparison across the physical sciences \cite{sharma2017mcmc,vonToussaint2011,trotta2008bayes}. However, its primary practical bottleneck is computational: likelihoods are expensive, parameter spaces are high-dimensional, and posteriors are often multi-modal. Consequently, the computational cost of an analysis is dictated by how efficiently a sampler extracts posterior information from likelihood evaluations.

Gravitational-wave (GW) astronomy exemplifies these challenges and serves as the benchmark setting for our method. Since LIGO's first detection in 2015 \cite{LIGOScientific:2016aoc}, characterizing astrophysical sources from noisy data has consistently relied on Bayesian methods \cite{Christensen:2022bxb,Thrane:2018qnx}, with evidence-based model selection used to resolve debates such as the nature of GW190521 \cite{LIGOScientific:2020ufj} and to test general relativity. These computational challenges are especially acute for current and next-generation GW observatories, whose overlapping signals, complex waveforms and global-fit analyses produce posteriors that are at once multi-modal and high-dimensional; a regime detailed in Sec.~\ref{sec:application}.

Current sampling methods, such as Markov Chain Monte Carlo (MCMC) and nested sampling, provide practical solutions for Bayesian inference but still face limitations when applied to realistic likelihood distributions. MCMC methods are widely used across the physical sciences, including GW astronomy, due to their ability to explore high-dimensional spaces efficiently \cite{Ashton_2021}, and can provide evidence estimation via techniques such as thermodynamic integration. However, they rely on local moves and suffer from poor mixing when likelihood distributions are highly multi-modal, leading to slow convergence \cite{latuszyński2025mcmcmultimodaldistributions}. To improve mixing, parallel tempering MCMC (PTMCMC) is typically employed, where multiple parallel chains with different temperatures interact during exploration \cite{Karnesis:2023ras,PhysRevLett.57.2607,doi:10.1143/JPSJ.65.1604,10.1093/mnras/stv2422}, but this significantly increases computational costs and the required degree of fine-tuning.

Nested sampling, though powerful for evidence estimation \cite{Buchner_2023,Skilling2006NestedSF,Higson_2018,Feroz:2009de,Albert:2024zsh,Buchner:2021cql} and in cross-disciplinary use from cosmology to materials science \cite{Ashton2022nested}, can also struggle with the multi-modal and heavy-tailed posteriors in GW analysis \cite{Dittmann_2024}. In multi-modal settings, the fundamental idea of shrinking the sampling region breaks down, as the region cannot be smaller than the separation between distinct modes \cite{Ashton:2022grj}. Accurate mode identification requires clustering of live points, but this is hindered by their limited number and noisy updates at each iteration, often resulting in unstable cluster assignments and unreliable allocation of live points \cite{Cai_2022,Feroz:2007kg,Shaw:2007jj}. If the live point allocation is suboptimal, entire modes may be underestimated or completely missed, compromising both standard and dynamic nested sampling \cite{dynesty2023}.
While increasing resolution and adopting tuning strategies (e.g., more live points, better clustering) can mitigate some of these issues, they come with a significant computational cost and require careful hyperparameter tuning, which limits the practical efficiency of nested sampling in high-dimensional GW problems.

Importance sampling (IS) provides a general approach for posterior evaluation and evidence estimation \cite{Kolmus:2021buf,Saleh:2024tgr,Wraith:2009if,Nitz:2024nhj,Dax:2022pxd,Kejriwal:2025upp}.
IS underpins methodologies well beyond GW astronomy; within GW analysis specifically, recent efforts include using IS to approximate high-fidelity posteriors from low-fidelity likelihood samples \cite{Saleh:2024tgr}, to leverage metric-based likelihood tiling \cite{Nitz:2024nhj}, and to incorporate neural networks for proposal adaptation \cite{Dax:2022pxd}, among others. Sequential Monte Carlo (an extension of IS) with temperature annealing has also been validated for GW inference, demonstrating robust performance through adaptive exploration of complex posterior landscapes \cite{williams2025validatingsequentialmontecarlo}. Importance sampling is likewise a possible step that simulation-based inference (SBI) uses to restore exactness to an approximate posterior learned from simulations \cite{papamakarios2016fast,cranmer2020frontier,Dax:2022pxd}. While highly effective when such networks can be successfully trained, training is not always feasible, which is why general-purpose samplers remain necessary; we return to the comparison in Sec.~\ref{sec:application}.

Traditional IS methods are most effective when the structure of the target distribution is well understood or hyperparameters are optimized, which is challenging for distributions where prior knowledge is limited. Adaptive importance sampling (AIS) improves IS by iteratively refining proposal distributions, which is particularly crucial in complex cases to reduce estimator variance \cite{bugallo2017adaptive,Paananen_2021,pal2023populationmontecarlonormalizing,Elvira_2019,ellaham2019efficient,cornuet2011adaptivemultipleimportancesampling,Bugallo2017,Paananen_2021,dalbey2014gaussian,li2015adaptive,Elvira_2022,martino2015adaptive}. However, AIS typically assumes a fixed proposal form, limiting its adaptability to complex likelihoods; it also struggles with large variance in importance weights \cite{korba2021adaptiveimportancesamplingmeets}, especially in noisy likelihoods. Some methods incorporate MCMC-assisted adaptation \cite{Mousavi_2021}, but this increases computational cost. Overcoming these challenges requires a more adaptive and efficient approach to proposal refinement.

\subsection{Executive summary of the method}
\label{subsec:execsum}
In this paper, we propose Parallel Adaptive Reweighting Importance Sampling (PARIS; descriptively, ``seed sampling'', the working name under which the method was first developed and which survives as a secondary name below), a novel scheme that integrates global exploration with local adaptation to achieve efficient sampling in complex target distributions. The algorithm begins by employing global Latin hypercube sampling (LHS) to identify high-probability points across the entire parameter space, which then serve as seeds to initialize independent Adaptive Reweighting Importance Sampling (ARIS) processes. Crucially, each ARIS process constructs its own mixture proposal (we adopt Gaussian mixtures in this paper), where the individual proposal components are defined by the samples generated within that process. Specifically, each component is centered on an individual sample, with its scale estimated from the weighted sample set, and its weight determined by the corresponding importance weight. As new samples are drawn from the updated proposal, they are incorporated into the process’ sample set, triggering a global reweighting of all samples (both new and old) according to the composite proposal; this weight update follows the standard multi-proposal IS framework by summing over all past proposals, thereby ensuring unbiased estimation across iterations \cite{Elvira_2019,Elvira_2015,he2014optimalmixtureweightsmultiple}. Moreover, by facilitating interactions among the independent AIS processes through merging those in close proximity based on a covariance-adjusted distance metric, the algorithm effectively avoids redundant exploration and enhances efficiency in multi-modal settings.

PARIS holds significant potential for astrophysical applications, offering several advantages over traditional approaches. 1) By leveraging LHS for initial space exploration, we ensure systematic and uniform coverage of the parameter space, reducing the risk of clumping in high-dimensional spaces. 2) Incorporating previous samples as component centers of a mixture proposal distribution allows it to flexibly adapt to complex, multimodal distributions without assuming a fixed distribution family. 3) PARIS dynamically adjusts importance weights: if a sample receives an excessively high weight, new samples are drawn around it, increasing the local proposal density. This reweighting mechanism automatically adjusts the level of exploration across different regions of parameter space, while a practical scheme is designed to make weight evaluation computationally efficient. 4) Each local adaptive process maintains its own sample set and mixture proposal, naturally forming clusters corresponding to distinct modes. When two processes become sufficiently close in distribution, they are merged to prevent redundant exploration. 5) Finally, method parameters are intuitively linked to prior knowledge, facilitating effective tuning. Mode scales estimate the density of LHS points, the number of modes guides the number of initial processes, and the merging threshold reflects mode separation assumptions.

The remainder of this paper is organized as follows: Sec.\ref{sec:background} provides background on IS and adaptive methods, laying the foundation for understanding the proposed approach. Sec.\ref{sec:Base Implement} introduces the base implementation of the PARIS algorithm, detailing its adaptive strategy and parallel processing framework. 
Sec.\ref{sec:Practicle Implement} presents the practical implementation of PARIS, incorporating optimizations for computational efficiency. Sec.\ref{sec:application} evaluates PARIS on a set of generic test problems and then on GW applications, comparing it to other modern samplers and demonstrating its effectiveness across diverse applications. Finally, Sec.\ref{sec:Discussion} discusses our results and suggests directions for future research. A summary of notation used throughout the paper is provided in Table~\ref{tab:notation} in Supplementary Information~\ref{app:notation}, although every symbol is also defined in the text where it first appears. The consistency proof, the covariance estimation scheme and the configurations of the astrophysical applications are likewise given in the Supplementary Information. An open-source Python implementation of the PARIS algorithm is publicly available at \url{https://github.com/mx-Liu123/parismc}.


\section{Background} \label{sec:background}

Given observational data \( D \), which is a vector in \(\mathbb{R}^{k}\), and a model \( M \), the goal is to infer the unknown parameters \( x \), which form a vector in \(\mathbb{R}^{p}\). The parameters \( x \) describe properties of the model \( M \), such as masses and spins in a binary black hole system. The posterior distribution of \( x \) is given by Bayes' theorem \cite{gelman2013bayesian}:
\begin{equation}
\begin{aligned}
&\widetilde{P}(x \mid D, M) = \frac{L(x \mid D, M) \, \pi(x \mid M)}{Z(D \mid M)} \\
&\propto P(x \mid D, M) = L(x \mid D, M) \, \pi(x \mid M).
\end{aligned}
\end{equation}

Here, the likelihood \( L(x \mid D, M) = P(D \mid x, M) \) measures data fit, and the prior \( \pi(x \mid M) \) encodes prior beliefs about the parameters. The evidence \( Z(D \mid M) \) normalizes the posterior but is typically intractable to compute because it requires integration over the high-dimensional parameter space. Consequently, we perform inference by sampling from the unnormalized posterior \( P(x \mid D, M) \) to approximate the posterior and its expectations without computing \( Z \).
Hence, the purpose of this paper is to develop a method for efficiently sampling in proportion to the unnormalized posterior \( P(x \mid D, M) \), although the method is applicable to arbitrary target distributions. For brevity, we refer to \( P(x \mid D, M) \) as \( P(x) \) and \( Z(D \mid M) \) as \( Z \) throughout the rest of the paper, and we write $p$ for the dimension of the parameter vector $x$.

\subsection{Importance Sampling}
Sampling from a tractable proposal rather than from the target is common to most Monte Carlo methods; what distinguishes IS is how this distributional mismatch is resolved. Rather than accepting or rejecting draws so as to leave the target invariant, IS keeps every draw and assigns each sample an importance weight \cite{robert2004monte}, thereby preserving sample independence and yielding the Bayesian evidence from the same run at no additional cost. Given a proposal distribution \( q(x) \), a sample is drawn from \( q(x) \) at every iteration, and its importance weight corrects for the discrepancy between \( q(x) \) and the target posterior \( P(x) \).
The importance weight associated with sample \( x_t \sim q(x) \) at iteration $t$ is defined as
\begin{equation}
w_t = \frac{P(x_t)}{q(x_t)}.
\end{equation}

Using these weights, after $T$ iterations,
the IS estimate of the normalized posterior distribution is given by
\begin{equation}
\hat{P}_{\text{IS}}(x) = \frac{1}{T \hat{Z}} \sum_{t=1}^{T} w_t \delta(x-x_t),
\end{equation}
where $T$ also represents the total number of samples drawn from the proposal distribution \( q(x) \), and \( \delta(x-x_t) \) is the Dirac delta function centered at sample \( x_t \) \cite{cornuet2011adaptivemultipleimportancesampling, marin2017consistency}. The term \( \hat{Z} \) is the estimated Bayesian evidence, computed as
\begin{equation}\label{eq:evidence}
\hat{Z} = \frac{1}{T} \sum_{t=1}^{T} w_t.
\end{equation}

This formulation shows that IS provides a weighted empirical approximation of the posterior, where the samples \( x_t \) are assigned weights to correct for the mismatch between the proposal and target distributions. Additionally, since the evidence \( Z \) is the integral of the unnormalized posterior over the parameter space, it can be directly estimated via the average importance weight.

\subsection{Multiple Proposal and Reweighting Strategy}

In cases where the proposal distribution varies across iterations, such that at each iteration $t$, a sample \( x_t \) is drawn from a different proposal distribution \( q_t(x) \),
the importance weights must be adjusted accordingly because the standard IS assumption—that all samples are drawn from a single fixed proposal distribution—no longer holds. Hence a ``deterministic-mixture'' reweighting strategy is employed, where previously drawn samples are retained and their weights are updated dynamically as new proposals are introduced \cite{cornuet2011adaptivemultipleimportancesampling}.

Under this strategy, the proposal distribution is treated as a mixture over all $q_t$, and the accumulated samples are treated as being drawn from this mixture. Specifically, instead of a single proposal distribution \( q(x) \), we maintain a cumulative mixture of proposal distributions across $T$ iterations. The importance weight of each sample \( x_t \) in sample set $\{x_t\}_{t=1}^{T}$ is therefore given by
\begin{equation}\label{eq:multprop}
w_t = \frac{P(x_t)}{\frac{1}{T} \sum_{t'=1}^{T} q_{t'}(x_t)},
\end{equation}
where the denominator represents the averaged mixture of all proposal distributions up to the current iteration, thus ensuring normalization. As a new sample is added, all existing sample weights are recomputed based on the updated mixture of proposals. By applying the reweighting strategy, previously drawn samples remain useful throughout the process, preventing the loss of valuable information.

\subsection{Adaptive Multiple Importance Sampling}

Adaptive Multiple Importance Sampling (AMIS) incorporates the sample reweighting strategy to improve efficiency \cite{cornuet2011adaptivemultipleimportancesampling,Bugallo2017}. In AMIS, the proposal itself is updated from previous samples.
At iteration $t$, a sample \( x_t \) is drawn from a parametric proposal distribution \( q(x \mid \theta_t) \) with parameters $\theta_t$, and all existing samples are reweighted; the importance weight for each sample \( x_t \) in the full sample set is thus given by Eq.~\eqref{eq:multprop} with $q_{t'}(x_t)\equiv q(x_t|\theta_{t'})$, a shorthand we use for the rest of the paper.

The parameters $\theta_t$ are updated iteratively from past weighted samples using some strategy $\Theta$: either (i) a chosen parametric family updated by moment matching (e.g., fit a Student's $t$-distribution by weighted mean and covariance) or by Kullback--Leibler (KL) divergence minimization (equivalently, weighted maximum likelihood within the family); or (ii) Gaussian mixture models (GMMs) refined by a weighted expectation-maximization (EM) update of mixture weights, means, and covariances. These choices are representative rather than exhaustive: any map $\Theta$ from the weighted samples to the proposal parameters is admissible, provided that the weights remain governed by Eq.~\eqref{eq:multprop}. The scheme developed below constitutes one such map, falling outside both categories.

{\footnotesize \singlespacing
\begin{algo}\label{algo:AMIS}{Adaptive Multiple Importance Sampling (AMIS)}

\vss \noindent \textbf{Input:}
Initial proposal parameters \(\theta_{1}\).

\vss \noindent \textbf{Iteration \((t=1,\dots,T)\)}
\begin{itemize}
  \item[1)] \textbf{Draw a new sample} \(x_{t}\sim q(x\mid\theta_{t})\).
  Let the accumulated sample set be \(\{x_{j}\}_{j=1}^{t}\).

  \item[2)] \textbf{Reweight all samples (deterministic–mixture weights):}
  \[
    w_j \;\leftarrow\; \frac{P(x_j)}{\displaystyle \frac{1}{t}\sum_{t'=1}^{t} q(x_j\mid\theta_{t'})}\,,\qquad j=1,\ldots,t.
  \]

  \item[3)] (Optional) \textbf{Compute estimated evidence and posterior:}
  \[
    \hat Z_{t} \;=\; \frac{1}{t}\sum_{j=1}^{t} w_j\,, \qquad
    \hat P_{t}(x) \;=\; \frac{1}{t\hat Z_{t}}\sum_{j=1}^{t} w_j\,\delta(x-x_j)\,.
  \]

  \item[4)] \textbf{Update proposal parameters:}
  \[
    \theta_{t+1}\ \leftarrow\ \Theta\!\Big(\{(x_j,w_j)\}_{j=1}^{t}\Big)\,.
  \]
\end{itemize}

\vss \noindent \textbf{Output:} Weighted samples \(\{(x_j,w_j)\}_{j=1}^{T}\), evidence \(\hat Z_T\), and the empirically recovered posterior \(\hat P_T\).

\vss \noindent \textit{Remark.}
Classical AMIS draws \(N_t\ge1\) samples per iteration and employs the general deterministic–mixture weight formulation \cite{cornuet2011adaptivemultipleimportancesampling}; here and in the base implementation of Sec.~\ref{sec:Base Implement} we take \(N_t=1\). The practical implementation of Sec.~\ref{sec:Practicle Implement} also retains one sample per iteration, but repeats a draw that falls outside the prior; those draws carry zero weight and are counted in \(N_t\), which is the general case above and is why \(N_t\) enters the denominator of the evidence estimate.
\end{algo}
}

The core of AMIS lies in the proper selection of an adaptive parametric proposal and an update strategy for its parameters. If chosen appropriately, AMIS can balance exploration and exploitation by dynamically refining the proposal distribution to better fit the target posterior. For AMIS and related methods such as population Monte Carlo \cite{pal2023populationmontecarlonormalizing,Elvira_2022}, the overall efficiency of sampling is influenced by the flexibility (or degrees of freedom) of the proposal, which allows it to better adapt to complex or multi-modal target distributions.

Standard AMIS proposal families and update rules have practical limitations: $t$-families are unimodal with elliptical symmetry; KL-based weighted MLE fits can be sensitive to misspecification; and GMMs require choosing the number of components while incurring extra cost from evaluating many components per sample per iteration. Moreover, naïve deterministic–mixture reweighting yields \(O(t^2)\) cost per process (each of the \(t\) stored samples is evaluated under all \(t\) proposals).
Our proposed algorithm PARIS is a variant of AMIS that mitigates both issues by adopting a sample-centered mixture proposal together with a sliding-window, rolling-denominator scheme whose cost scales linearly with the window size while preserving estimator quality.

\section{PROPOSED ALGORITHM: PARIS (Base Implementation)}\label{sec:Base Implement}

The rationale behind this construction, and how its hyperparameters are informed by prior knowledge, are outlined in the overview of Sec.~\ref{subsec:execsum}. In this section, we present an idealized ``base'' version of the PARIS algorithm to illustrate the core ideas behind it. This version is designed to clearly demonstrate the adaptive strategy and parallel structure, while keeping the implementation minimal and transparent. A more numerically stable and computationally efficient version of the algorithm will be introduced in Sec.~\ref{sec:Practicle Implement}. The asymptotic consistency of the estimator constructed from the PARIS algorithm is addressed in Supplementary Information~\ref{app:convergence}.

\subsection{Adaptive Reweighting Strategy}\label{subsec:Adaptive_Strategy}
We adopt a sample-centered mixture as the proposal: each stored sample serves as a Gaussian component whose mean is fixed at that sample; all components share a covariance estimated from the current weighted set and updated periodically; and component weights are the normalized importance weights. This differs from a conventional EM-fitted finite Gaussian mixture: we do not run EM, we do not preselect a fixed number of components, and we avoid per-component covariance estimation. The mixture size increases automatically with the number of stored samples, removing model-order selection and reducing sensitivity to initialization.

We now define the construction of an ARIS process that utilizes such an adaptive proposal. Suppose the current iteration is $t$. Given the sample set
\begin{equation}
\{ x_{t'} \}_{t'=1}^{t-1},
\end{equation}
with corresponding importance weights
\begin{equation}
\{ w_{t'} \}_{t'=1}^{t-1},
\end{equation}
we define the normalization
\begin{equation}
Z_{t-1} \;=\; \sum_{t'=1}^{t-1} w_{t'}\,,
\end{equation}
as well as the weighted covariance (using samples up to iteration $t-1$),
\begin{equation}
\label{eq:cov}
\Sigma_{t-1} \;=\; \frac{1}{Z_{t-1}}\sum_{t'=1}^{t-1} w_{t'} \,\big(x_{t'} - \mu_{t-1}\big)\big(x_{t'} - \mu_{t-1}\big)^\top \, ,
\end{equation}
where the weighted mean vector is
\begin{equation}
\mu_{t-1} \;=\; \frac{1}{Z_{t-1}}\sum_{t'=1}^{t-1} w_{t'}\, x_{t'} \,.
\end{equation}
A Gaussian mixture proposal at iteration $t$ can then be explicitly defined as:
\begin{equation}\label{eq:mixture_density}
q_{t}(x) \;=\; \frac{1}{Z_{t-1}}\sum_{t'=1}^{t-1} w_{t'}\, \mathcal{N}\!\left(x \,\middle|\, x_{t'}, \Sigma_{t-1}\right) \,,
\end{equation}
where $Z_{t-1}$ serves as a normalization factor. Following Eq.~\eqref{eq:evidence}, the corresponding evidence estimate is the mean weight, $\hat{Z}_{t-1} = Z_{t-1}/(t-1)$; we reserve the hat for estimates of $Z$ and write $Z_{t-1}$ for the weight sum throughout. To draw a new sample \( x_{t} \) from this adaptive proposal:
1) Choose a sample \( y_{t} \in \{ x_{t'} \}_{t'=1}^{t-1} \) according to their normalized weights;
2) Sample \( x_{t} \sim \mathcal{N}(x \mid y_{t}, \Sigma_{t}) \).

Over \( t \) iterations, the cumulative proposal distribution (i.e., the mixture over all previous proposals) will be updated as:
\begin{equation}
q_{\text{total}}(x) \leftarrow \frac{1}{t} \sum_{t'=1}^{t} q_{t'}(x),
\end{equation}
where the factor \( \frac{1}{t} \) ensures normalization across iterations. Thus, the importance weights of all collected samples \( \{ x_{t'} \}_{t'=1}^{t} \) can be updated as:
\begin{equation}
w_{t'} \leftarrow \frac{P(x_{t'})}{q_{\text{total}}(x_{t'})}, \quad \text{for } t' = 1, \ldots, t,
\end{equation}
where \( P(x_{t'}) \) is the target posterior density.

The adaptive behavior of the proposal distribution arises directly from the cumulative reweighting process:
as the overall proposal \(q_{\text{total}}\) evolves with each iteration, samples drawn from regions where it remains small naturally receive higher importance weights.
Subsequent proposals—constructed from the weighted sample set—therefore allocate more probability mass near these regions.
As \(q_{\text{total}}\) increases locally with continued sampling, the corresponding weights shrink, producing a self-stabilizing adaptation that automatically balances exploration and exploitation without any manual tuning or heuristic correction.

\subsection{Parallel Processes}

One of the key advantages of PARIS lies in its inherent parallel structure, which enables multiple sampling processes to run concurrently and almost independently. A notable computational benefit of this approximate independence is that the importance weight for each sample need not account for the mixture proposals of all processes. With $j$ as the process index, we have in principle:
\begin{equation}
w_{t}^{(j)} \;=\;
\frac{P(x_{t}^{(j)})}{\frac{1}{N_\text{proc}}\sum_{j'=1}^{N_\text{proc}} q_{\text{total}}^{(j')}(x_{t}^{(j)})}\,,
\label{eq:global-mixture-denom}
\end{equation}
where $N_\text{proc}$ is the current number of processes. However, if different processes
explore non-overlapping or only weakly overlapping regions (maintained via the merging
criterion described below), we may safely approximate
\begin{equation}
w_{t}^{(j)} \;\approx\; \frac{P(x_{t}^{(j)})}{\frac{1}{N_\text{proc}}\textbf{}q_{\text{total}}^{(j)}(x_{t}^{(j)})}\,,
\label{eq:local-denominator}
\end{equation}
effectively treating each process as locally self-contained. The validity of this approximation for any given process $j$ relies on the assumption that in the denominator of Eq.~\eqref{eq:global-mixture-denom}, the contributions from other processes are negligible compared to the self-term; that is, $\sum_{j'\neq j} q_{\text{total}}^{(j')}(x_{t}^{(j)}) \ll q_{\text{total}}^{(j)}(x_{t}^{(j)})$. To enforce this condition, we introduce a merging criterion based directly on the relative magnitude of these proposal densities.

For any two active processes $j$ and $j'$ at iteration $t$, we assess whether process $j'$ has significantly encroached upon the region explored by process $j$. A conflict arises if the ``foreign'' proposal density $q_{\text{total}}^{(j')}$ exceeds the ``local'' proposal density $q_{\text{total}}^{(j)}$ at any of $j$'s historical samples. Formally, this occurs when:
\begin{equation}
\exists t' \in \{1,\dots,t\} \quad \text{such that} \quad q_{\text{total}}^{(j')}(x_{t'}^{(j)}) > q_{\text{total}}^{(j)}(x_{t'}^{(j)}) \,.
\end{equation}
When this condition is met, the term $q_{\text{total}}^{(j')}(x_{t'}^{(j)})$ becomes a dominant component in the global mixture denominator for sample $x_{t'}^{(j)}$, violating the independence assumption. Consequently, the two processes are identified as exploring the same mode and must be merged.

In practice, at each iteration, we collect any pairs of processes that satisfy the above condition into clusters. Within each cluster, we retain only the process with the highest posterior peak found so far and terminate the others. This ensures that the remaining active processes are sufficiently distinct, justifying the use of the computationally efficient local weight approximation \eqref{eq:local-denominator}. Such a density-based clustering scheme effectively identifies and eliminates redundant processes exploring the same posterior mode, thereby preventing the overcounting of local evidence contributions.

Since the process retained in each cluster is selected by peak height rather than integrated posterior mass, a natural concern is whether merging biases the evidence or mode weights in challenging scenarios: nearby modes, ridges, unequal-volume modes, strong correlations, bridges, or secondary modes lying only marginally below the primary peak. Two properties mitigate this risk. First, our open-source implementation retains the samples of a terminated process along with their weights, ensuring they remain available to downstream estimators; merging thus eliminates redundant future exploration rather than discarding accumulated information. Second, because the surviving process constructs a mixture proposal centred on its recent samples, its support naturally covers their local neighbourhood, a region already verified by the merge criterion to encompass the parameter domain of the terminated process. Section~\ref{subsec:xshape} comprehensively evaluates these scenarios on a benchmark target with analytically tractable evidence and mode weights.

Finally, a suitable initialization scheme is needed to seed a set of parallel processes for effective global exploration. In PARIS, the initial locations of processes are determined using LHS, a type of quasi-Monte Carlo (QMC) method designed to ensure better coverage of the parameter space than purely random sampling \cite{saves2024smt,SMT2019}. LHS stratifies each dimension of the space into equally probable intervals and samples within each interval, promoting space-filling properties and reducing sample clustering. Depending on the task, LHS might also be replaced by other QMC methods \cite{Dick_Kuo_Sloan_2013}. Orthogonal Latin hypercube designs \cite{tang1993orthogonal,owen1992orthogonal} are one such option and could place seeds closer to the modes; as our concern here is the method itself, we take the plain construction as a baseline.

{\footnotesize \singlespacing
\begin{algo}\label{algo:simple_implement}{PARIS (Base Implementation)}

\vss \noindent \textbf{Input:}
\begin{itemize}
    \item Distribute \( N_{\text{LHS}} \) LHS points over the prior and select \( N_{\text{seed}} \) points with the highest posterior as seeds \( \{x_{\text{seed}}^{(j)}\}_{j=1}^{N_{\text{seed}}} \).
    \item Initialize the set of active processes \(\mathcal{J} = \{1, \dots, N_{\text{seed}}\}\) and the initial proposal covariance \(\Sigma_{1}^{(j)}=\Sigma_\mathrm{init}\).
\end{itemize}

\vss \noindent \textbf{Iteration \((t = 1,\dots,T)\)}
\begin{itemize}
    \item[\textbf{1)}] \textbf{Parallel Sampling \& Weighting:} For each active process \( j \in \mathcal{J} \):
    \begin{itemize}
        \item[\textbf{1.1)}] \textbf{Construct Proposal \( q_{t}^{(j)} \):}
        \begin{itemize}
            \item If \( t=1 \): Set \( q_{1}^{(j)}(x) = \mathcal{N}(x \mid x_{\text{seed}}^{(j)}, \Sigma_{1}^{(j)}) \).
            \item If \( t>1 \): Compute weighted covariance \(\Sigma_{t}^{(j)}\) from past samples \(1,\dots,t-1\). Define the mixture:
            \[
            q_{t}^{(j)}(x) \;\propto\; \sum_{t'=1}^{t-1} \, w_{t'}^{(j)} \,\mathcal{N}\bigl(x \mid x_{t'}^{(j)}, \,\Sigma_{t}^{(j)}\bigr).
            \]
        \end{itemize}

        \item[\textbf{1.2)}] \textbf{Draw Sample:}
        Draw \( x_t^{(j)} \sim q_{t}^{(j)}(x) \) (by first selecting a component index \(t' \in \{1,\dots,t-1\}\) proportional to \(w_{t'}^{(j)}\), then sampling from that Gaussian).

        \item[\textbf{1.3)}] \textbf{AMIS Weight Update:}
        Update weights for all samples \( t' \in \{1, \dots, t\} \):
        \[
        w_{t'}^{(j)} \longleftarrow \frac{P(x_{t'}^{(j)})}{\displaystyle \frac{1}{t}\sum_{t''=1}^t q_{t''}^{(j)}(x_{t'}^{(j)})}\,.
        \]

        \item[\textbf{1.4)}] \textbf{Storage:} Update trajectory data \[\{(x_{t'}^{(j)},w_{t'}^{(j)},q_{t'}^{(j)})\}_{t'=1}^{t}\,.\]

    \end{itemize}

        \item[\textbf{2)}] \textbf{Process Interaction (Merging):}
    \begin{itemize}
        \item[\textbf{2.1)}] \textbf{Check Dominance:} For every pair of active processes $j$ and $j'$, check if $j'$ dominates $j$ at any sample point:
        \[
             \exists t' \in \{1,\dots,t\} \text{ s.t. } q_{\text{total}}^{(j')}(x_{t'}^{(j)}) > q_{\text{total}}^{(j)}(x_{t'}^{(j)})\,.
        \]
        \item[\textbf{2.2)}] \textbf{Clustering:} Construct a directed graph where an edge $j' \to j$ exists if the above is true. Identify weakly connected components as clusters.
        \item[\textbf{2.3)}] \textbf{Selection:} In each cluster, retain the process with the highest posterior peak. Terminate others and update the active set $\mathcal{J}$.
    \end{itemize}
    \item[\textbf{3)}] \textbf{Evidence Update:}
    \[
        \hat{Z}_{t} = \frac{1}{N_\mathrm{proc}\,t} \sum_{j' \in \mathcal{J}}\sum_{t'=1}^{t} w_{t'}^{(j')}\,,
    \]
    where $N_\mathrm{proc}$ is the number of active processes, so that $N_\mathrm{proc}\,t$ is the total number of proposal draws made.
\end{itemize}

\end{algo}
}

\section{PROPOSED ALGORITHM: PARIS (Practical Implementation)}\label{sec:Practicle Implement}

In this section, we present the practical implementation of the PARIS algorithm, which incorporates additional strategies for improved computational efficiency and numerical stability. This is the version used in the experiments reported in the following section. To standardize the sampling space, all priors of the target distribution's parameters are normalized (re-scaled) to lie within the $[0,1]$ interval.

\subsection{Covariance Update Strategy}
In our adaptive strategy, the covariance matrix is updated only once every $\gamma$ iterations. Thus, the notation for the process covariance at iteration $t$ becomes \(\Sigma_{\left\lceil t/\gamma \right\rceil}\), where \(\left\lceil \cdot \right\rceil\) denotes the ceiling operator. This design choice significantly reduces memory usage by avoiding the need to store the full collection of historical covariances. More importantly, it allows us to employ more robust yet computationally expensive covariance estimation methods without introducing excessive overhead. For example, details on how we estimate the covariance matrix from samples truncated by the prior region—as well as how we apply conservative corrections when the number of samples is small—are provided in Supplementary Information~\ref{app:Cov}. The parameter \( \gamma \) also controls the initial iterations where the initial covariance is used, ensuring sufficient data accumulation before updating. If \( \gamma \) is too small, the new covariance, estimated from a limited sample size, may be unreliable and ad-hoc.

\subsection{Modified Gaussian for Proposal Regularization}

\begin{figure}[tbp]
    \centering
    \includegraphics[width=1\textwidth]{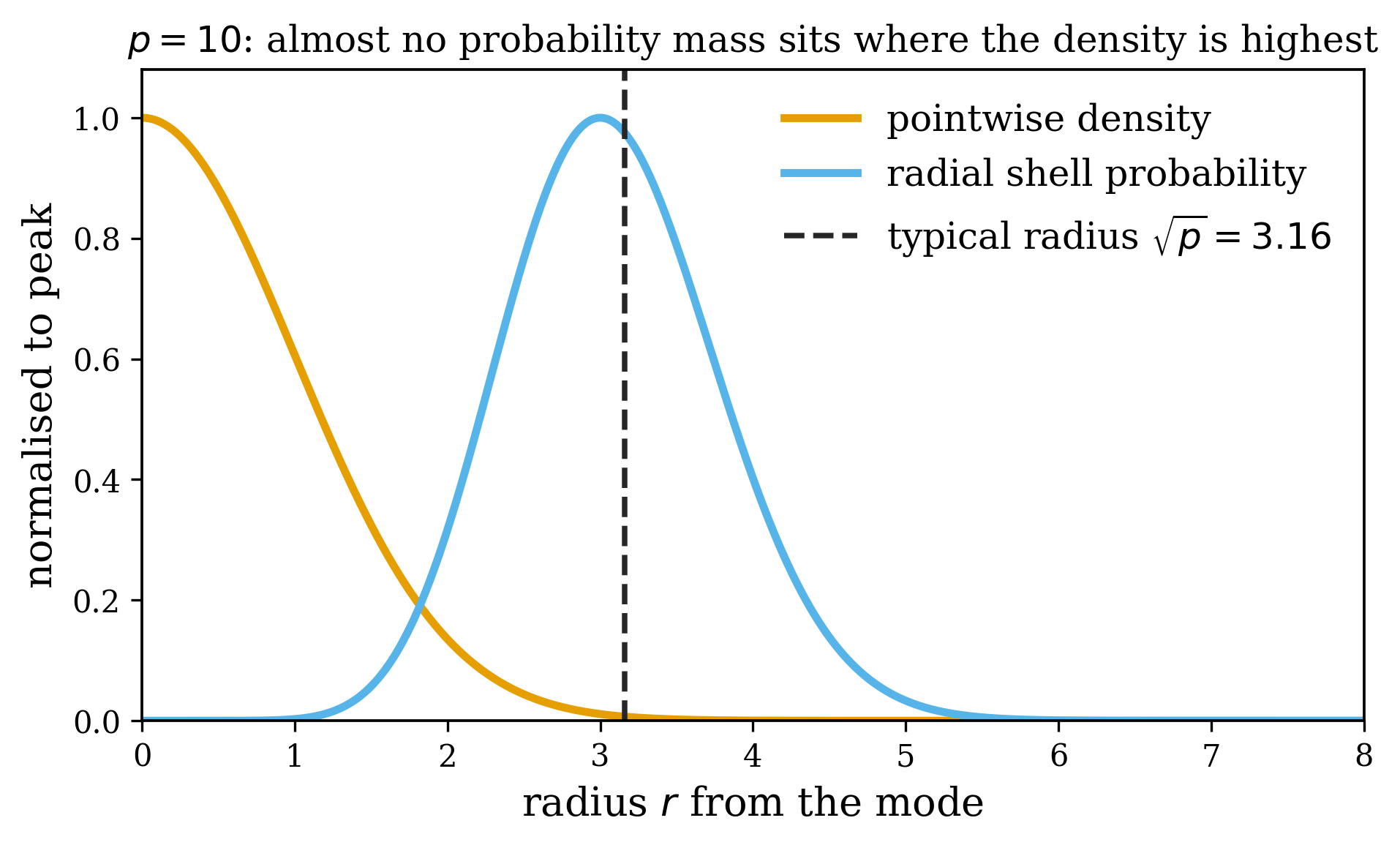}
    \caption{Pointwise density (orange) and radial shell probability (blue) for a
    10-dimensional Gaussian. Most probability mass lies near the typical radius
    \(\sqrt{p}\) rather than at the peak, so the self-term’s peak value overestimates
    local density by an exponential factor.}
    \label{fig:typicalset}
\end{figure}

In high dimensions, a $p$-dimensional Gaussian places most of its probability mass on a thin typical set at squared Mahalanobis radius approximately \(p\) (see Fig.~\ref{fig:typicalset}). However, for a kernel
\(\mathcal{N}(x\mid \mu,\Sigma)\propto \exp\!\big(-\tfrac12\|x-\mu\|_{\Sigma}^2\big)\),
where \(\|v\|_{\Sigma} := \sqrt{v^\top \Sigma^{-1} v}\) denotes the Mahalanobis distance,
the pointwise density at the mean \(\mu\) exceeds the typical pointwise density
(at \(\|x-\mu\|_{\Sigma}=\sqrt{p}\)) by a ``peak-to-typical'' factor
\[
\frac{\mathcal{N}(\mu\mid \mu,\Sigma)}{\mathcal{N}(x\mid \mu,\Sigma)}
\bigg|_{\|x-\mu\|_{\Sigma}=\sqrt{p}}
=\exp\!\Big(\frac{p}{2}\Big).
\]
In other words, most samples from high-dimensional Gaussians will lie far from the mean, even though the probability density at their locations is much lower.

Under deterministic--mixture reweighting, when a new sample \(x_t\) is also used as the
next component center \(y_t=x_t\), the denominator at \(x_t\)
includes a self-term evaluated at its own mean,
\(\mathcal{N}(x_t\mid y_t,\Sigma)=(2\pi)^{-p/2}|\Sigma|^{-1/2}\).
This self-term can dominate the mixture denominator,
causing the importance weight \(w_t\) to be artificially small even when \(x_t\)
reveals a new mode. To correct for this bias, which we term ``self-density inflation'',
we calibrate the self-term down to the density value on the typical shell.
We define a modified Gaussian function, denoted by \( \widetilde{\mathcal{N}} \), whose peak value at the mean
is rescaled. Specifically, instead of the usual value at the mean
\begin{equation}
\mathcal{N}(\mu \mid \mu, \Sigma) = \frac{1}{(2\pi)^{p/2} |\Sigma|^{1/2}},
\end{equation}
we define
\begin{equation}
\widetilde{\mathcal{N}}(\mu \mid \mu, \Sigma)
= \frac{1}{(2\pi)^{p/2} |\Sigma|^{1/2}} \exp\!\left(-\frac{p}{2}\right).
\end{equation}

The factor \(\exp(-p/2)\) cancels out the peak-to-typical factor of a \(p\)-dimensional Gaussian, thereby calibrating the self-term to the desired value.
In our implementation, the modified function \(\widetilde{\mathcal{N}}\) is used consistently both during exploration and in the final deterministic–mixture weights, because the self-term otherwise dominates the mixture denominator even at convergence.
The adjustment effectively restores balance between local self-density
and cross-component contributions, and leads to more robust performance overall.

\subsection{Various Approximations for Weight Evaluation}\label{subsec:weightapprox}

The computational burden of weight evaluation in the base implementation is substantial. At iteration $t$, for a generic sample \(x\), the cost of evaluating the cumulative proposal density up to that iteration,
\begin{equation}
\frac{1}{t} \sum_{t'=1}^{t} q_{t'}(x),
\end{equation}
is \( O(t^2) \) (since there are \(t(t+1)/2\) Gaussian components in the sum). Consequently, evaluating this density across all \(t\) samples incurs a total cost of \( O(t^3) \). To maintain scalability, we introduce two key approximations for the deterministic-mixture denominator:

\begin{equation}
\label{eq:propapprox}
\frac{1}{t} \sum_{t'=1}^{t} q_{t'} \approx q_t \approx \frac{1}{t} \sum_{t'=1}^{t} \widetilde{\mathcal{N}}(\cdot \mid y_{t'}, \Sigma_{\left\lceil t'/\gamma \right\rceil}).
\end{equation}

The first approximation is justified by the convergence of the evidence estimate $\hat{Z}_{t} = Z_{t}/t$, where $Z_{t} = \sum_{t'=1}^{t} w_{t'}$ is the cumulative weight sum, to a stable evidence value $Z$, which is both observed empirically and implied by the asymptotic consistency argued in Supplementary Information~\ref{app:convergence}. In other words, the individual new weights satisfy $w_{t}/Z_{t}\to0$ as $t \to \infty$, which also means that the sequence of proposals $\{q_{t'}\}$ becomes stationary. This allows the latest proposal $q_{t}$ to effectively represent the cumulative average of all previous proposals. Such an assumption is also made in, e.g., \cite{ellaham2019efficient}.



For the second approximation, we substitute the weighted mixture with an unweighted empirical sum, i.e.,
\begin{equation}q_t(x) \approx \sum_{t'=1}^{t} \frac{w_{t'}}{Z_{t}} \widetilde{\mathcal{N}}(x \mid x_{t'}, \Sigma_{t}) \approx \frac{1}{t} \sum_{t'=1}^{t} \widetilde{\mathcal{N}}(x \mid y_{t'}, \Sigma_{\lceil t'/\gamma \rceil}),
\end{equation}
where $\Sigma_{\lceil t'/\gamma \rceil} \approx \Sigma_t$ due to the convergence of the sample covariance estimator.
This substitution is valid since $y_{t'} \sim \sum \frac{w_{t'}}{Z_t} \delta(x-x_{t'})$, and the law of large numbers implies that for any test function $g$:
\begin{equation} \frac{1}{t} \sum_{t'=1}^{t} g(y_{t'}) \xrightarrow{t \to \infty} \sum_{t'=1}^{t} \frac{w_{t'}}{Z_{t}} g(x_{t'}).
\end{equation}


Building on these approximations, we introduce a sliding window over the most recent \( \alpha \) iterations. These latest \( \alpha \) samples are considered ``live'' samples, and only their importance weights are updated using:
\begin{equation}\label{eq:windowweights}
\begin{split}
w_{t'} &= \frac{P(x_{t'})}
{\frac{1}{\alpha} \sum_{t''=t-\alpha+1}^{t}
\widetilde{\mathcal{N}}(x_{t'} \mid y_{t''}, \Sigma_{\left\lceil t''/\gamma \right\rceil})}, \\
&\quad \text{for } t' = t - \alpha + 1, \dots, t.
\end{split}
\end{equation}
Earlier samples are treated as ``dead'' samples, and their weights are no longer updated. The hyperparameter \( \alpha \) controls the approximation precision.

Now, if we define the unnormalized denominator of importance weights:
\begin{equation}
d_{t'} = \sum_{t''=t - \alpha + 1}^{t}
\widetilde{\mathcal{N}}(x_{t'} \mid y_{t''}, \Sigma_{\left\lceil t''/\gamma \right\rceil}),
\end{equation}
then Eq.~\eqref{eq:windowweights} becomes
\begin{equation}
w_{t'} = \frac{P(x_{t'})}{\frac{1}{\alpha} d_{t'}}.
\end{equation}
At iteration $t$, the values of $d_{t'}$ for each of the \( \alpha \) samples can then be computed efficiently:
\begin{itemize}
    \item For the new sample \( x_t \), compute:
    \begin{equation}\label{eq:weight_new_sample}
    d_t = \sum_{t'=t - \alpha + 1}^{t}
    \widetilde{\mathcal{N}}(x_t \mid y_{t'}, \Sigma_{\left\lceil t'/\gamma \right\rceil}).
    \end{equation}

    \item For existing samples \( x_{t'} \), where \( t' = t - \alpha + 1, \dots, t - 1 \), update:
    \begin{align}
    d_{t'} \leftarrow &\, d_{t'} +
    \widetilde{\mathcal{N}}(x_{t'} \mid y_{t}, \Sigma_{\left\lceil t/\gamma \right\rceil}) \nonumber \\
    & - \widetilde{\mathcal{N}}(x_{t'} \mid y_{t - \alpha}, \Sigma_{\left\lceil (t-\alpha)/\gamma \right\rceil}).
    \end{align}

\end{itemize}

This rolling update efficiently incorporates the latest proposal while removing the oldest (now outside the live window), and avoids recomputing the full sum from scratch. As a result, the computational complexity per iteration is \( O(\alpha) \) rather than \( O(\alpha^2) \). Furthermore, discarding early proposals enhances the stability of importance weights within the live window. During the initial adaptive phase, the proposals evolve rapidly; removing these outdated components prevents the estimator from being biased or destabilized by high-variance weights associated with suboptimal early explorations.

While population Monte Carlo methods also utilize Gaussian mixture proposals, the number of mixture components (essentially $\alpha$ in our method) is typically limited to less than 100 due to computational constraints \cite{Elvira_2022,martino2015adaptive}. In contrast, our approach (selecting previous samples as component centers and leveraging a scalable weight update) enables the use of $\sim10^4$ components without prohibitive computational cost. This scalability allows the mixture to more closely approximate complex target distributions. In this work, we set the hyperparameter $\alpha$ to be 1000 for all of the experiments.

\subsection{Efficient Process Interaction}

The merging criterion introduced in the base implementation requires evaluating the full cumulative mixture densities for all historical samples. Specifically, evaluating the proposal density $q_{\text{total}}^{(j)}(x)$ at a single point involves summing over $t$ past iterations, where the proposal at each iteration is itself a mixture of up to $t$ components. This results in a computational cost of $O(t^2)$ per point evaluation. Since the original merging criterion requires verifying this condition across all $t$ historical samples for every pair of processes, the total complexity scales as $O(t^3)$. As $t$ increases, this cubic scaling becomes computationally prohibitive.

To maintain scalability, we propose a practical merging criterion that leverages the sliding window approximation in Sec. \ref{subsec:weightapprox}, and further restricts the check to the most recent sample $x_t^{(j)}$ from each process.
Recall that for each process $j$, the unnormalized proposal density at $x_t^{(j)}$ is approximated by Eq.~\eqref{eq:weight_new_sample}:
\begin{equation}
d_t^{(j)} = \sum_{t'=t - \alpha + 1}^{t}
\widetilde{\mathcal{N}}(x_t^{(j)} \mid y_{t'}^{(j)}, \Sigma_{\left\lceil t'/\gamma \right\rceil}^{(j)}).
\end{equation}
To test if a foreign process $j'$ overlaps with process $j$, we evaluate the unnormalized proposal density for $j'$ at $x_t^{(j)}$:
\begin{equation}\label{eq:cross_density}
d_t^{(j' \to j)} = \sum_{t'=t - \alpha + 1}^{t}
\widetilde{\mathcal{N}}(x_t^{(j)} \mid y_{t'}^{(j')}, \Sigma_{\left\lceil t'/\gamma \right\rceil}^{(j')}).
\end{equation}

As in the base implementation, a merger is triggered if the foreign process provides a higher density contribution than the local process, but here the check is weakened to only the current sample point:
\begin{equation}
d_t^{(j' \to j)} > d_t^{(j)}.
\end{equation}
In practice, this means that 1) ARIS processes will not merge as quickly as in the base implementation, and 2) the weight approximation \eqref{eq:local-denominator} is less valid during the exploration phase, but the impact of this is transitory and largely negligible in an adaptive algorithm.
More crucially, the cost of the merging criterion now also scales linearly with the window size $\alpha$ rather than cubically with the iteration count $t$, ensuring the algorithm remains efficient over long runs.

The check remains pairwise, so its per-sweep cost scales as $\mathcal{O}(N_\text{proc}^2)$, raising potential concerns about its scalability when numerous overlapping sources sustain many concurrent processes. Crucially, each pairwise test queries only stored proposal parameters and entirely avoids likelihood evaluations: at the dimensionality of the Galactic-binary search of Sec.~\ref{sec:application} a full sweep costs $0.7$ to $39$~ms for $N_\text{proc}$ between $50$ and $400$, against $10.8$~ms for one likelihood evaluation there (both timed on AMD Ryzen 9 9900X3D), so even a run permanently at $400$ processes would spend about $0.1\%$ of its wall clock here.

{\footnotesize \singlespacing
\begin{algo}\label{algo:practical_implement}{PARIS (Practical Implementation)}

\vss \noindent \textbf{Input:}
\begin{itemize}
    \item Distribute \( N_{\text{LHS}} \) LHS points over the prior and select \( N_{\text{seed}} \) points with the highest posterior as seeds \( \{x_{\text{seed}}^{(j)}\}_{j=1}^{N_{\text{seed}}} \).
    \item Parameters: Initial covariance \(\Sigma_{1}^{(j)}=\Sigma_\mathrm{init}\), sliding window size \(\alpha\), covariance update interval \(\gamma\).
    \item Initialize active processes set \(\mathcal{J} = \{1, \dots, N_{\text{seed}}\}\).
\end{itemize}

\vss \noindent \textbf{Iteration \((t = 1,\dots,T)\)}
\begin{itemize}
    \item[\textbf{1)}] \textbf{Proposal Adaptation \& Sampling:} For each \( j \in \mathcal{J} \):
    \begin{itemize}
        \item[\textbf{1.1)}] \textbf{Construct Proposal \( q_{t}^{(j)} \):}
        \begin{itemize}
            \item If \( t=1 \): \( q_{1}^{(j)}(x) = \widetilde{\mathcal{N}}(x \mid x_{\text{seed}}^{(j)}, \Sigma_{1}^{(j)}) \).
            \item \textbf{If} \( t > 1 \):
                \begin{enumerate}[label=\arabic*., leftmargin=2em]
                    \item \textbf{Covariance Update}: If \( (t-1) \bmod \gamma = 0 \), estimate new covariance \(\Sigma_{\lceil t/\gamma \rceil}^{(j)}\) using past weighted samples (see Supplementary Information~\ref{app:Cov}); otherwise, maintain the previous covariance.

                    \item \textbf{Mixture Construction}: Define the proposal as a mixture over the sliding window \( \mathcal{T}_{\text{win}} = \{ \max(1, t-\alpha), \dots, t-1 \} \):
                    \[
                    q_{t}^{(j)}(x) \propto \sum_{t' \in \mathcal{T}_{\text{win}}} w_{t'}^{(j)} \widetilde{\mathcal{N}}\left(x \mid x_{t'}^{(j)}, \Sigma_{\lceil t/\gamma \rceil}^{(j)}\right).
                    \]
                \end{enumerate}
        \end{itemize}

        \item[\textbf{1.2)}] \textbf{Draw Sample:}  Repeat steps (a)--(b) until acceptance:
        \begin{itemize}
            \item[\textbf{(a)}] Draw \( \widetilde{x}_{t}^{(j)} \sim q_{t}^{(j)}(x) \) (by first selecting a component index \(t' \in \{1,\dots,t-1\}\) proportional to \(w_{t'}^{(j)}\), then sampling from that Gaussian.

            \item[\textbf{(b)}] If \( \widetilde{x}_{t}^{(j)} \) is within the prior region or reach maximum number (set to 1000 in this work) of trials, set \( x_{t}^{(j)} \leftarrow \widetilde{x}_{t}^{(j)} \) and record trial number \( N_{t}^{(j)} \).
        \end{itemize}
    \end{itemize}

    \item[\textbf{2)}] \textbf{Sliding Window Weight Update:} For each \( j \in \mathcal{J} \):
    \begin{itemize}
        \item[\textbf{2.1)}] \textbf{Update Past Denominators:}
        For samples \( t' \) in the window \( \max(1, t-\alpha) \leq t' \leq t-1 \):
        \[
        d_{t'}^{(j)} \leftarrow d_{t'}^{(j)} + \widetilde{\mathcal{N}}(x_{t'}^{(j)} \mid y_{t}^{(j)}, \Sigma_{\lceil t/\gamma \rceil}^{(j)})\,.
        \]
        If \( t > \alpha \), remove the leaving component:
        \[
        d_{t'}^{(j)} \leftarrow d_{t'}^{(j)} - \widetilde{\mathcal{N}}(x_{t'}^{(j)} \mid y_{t - \alpha}^{(j)}, \Sigma_{\lceil (t - \alpha)/\gamma \rceil}^{(j)})\,.
        \]

        \item[\textbf{2.2)}] \textbf{Initialize New Denominator:}
        Compute for the new sample \( x_t^{(j)} \):
        \[
        d_t^{(j)} = \sum_{t'=\max(1,t-\alpha)}^{t} \widetilde{\mathcal{N}}(x_t^{(j)} \mid y_{t'}^{(j)}, \Sigma_{\lceil t'/\gamma \rceil}^{(j)})\,.
        \]

        \item[\textbf{2.3)}] \textbf{Compute Weights:}
        For all active samples \( t' \):
        \[
        w_{t'}^{(j)} \leftarrow \frac{P(x_{t'}^{(j)})}{\frac{1}{\min(t, \alpha)} \, d_{t'}^{(j)}}\,.
        \]

        \item[\textbf{2.4)}] \textbf{Storage:} Update trajectory data
        \[\{x_{t'}^{(j)}, P(x_{t'}^{(j)}), y_{t'}^{(j)}, d_{t'}^{(j)}, \Sigma_{\lceil t'/\gamma \rceil}^{(j)}\}_{t'=1}^{t}\,.\]

    \end{itemize}

        \item[\textbf{3)}] \textbf{Process Interaction (Merging):}
    \begin{itemize}
        \item[\textbf{3.1)}] \textbf{Check Dominance:} For every pair of active processes $j$ and $j'$, check if $j'$ dominates $j$ at the current sample:
        \[
        d_t^{(j' \to j)} = \sum_{t'=t - \alpha + 1}^{t}
        \widetilde{\mathcal{N}}(x_t^{(j)} \mid y_{t'}^{(j')}, \Sigma_{\lceil t'/\gamma \rceil}^{(j')}).
        \]
        Identify if $j'$ covers $j$ using the condition:
        \[
        d_{t}^{(j' \to j)} > d_{t}^{(j)}\,.
        \]

        \item[\textbf{3.2)}] \textbf{Clustering:} Construct a directed graph where an edge $j' \to j$ exists if the above is true. Identify weakly connected components as clusters.

        \item[\textbf{3.3)}] \textbf{Selection:} In each cluster, retain the process with the highest posterior peak. Terminate others and update the active set $\mathcal{J}$.
    \end{itemize}

    \item[\textbf{4)}] \textbf{Evidence Update:}
    \[
        \hat{Z}_{t} = \frac{\sum_{j' \in \mathcal{J}}\sum_{t'=t/2}^{t} w_{t'}^{(j')}}{\sum_{j' \in \mathcal{J}}\sum_{t'=t/2}^{t}N_{t'}^{(j')}}\,,
    \]
    where $N_{t'}^{(j')}$ is the trial number recorded in step 1.2(b), so the denominator is the total number of proposal draws; only the latest half of the samples is used.
\end{itemize}
\end{algo}
}

\section{Application to Toy Models and GW Analysis}
\label{sec:application}

In this section, we apply PARIS to various numerical experiments, and compare its performance to two other modern samplers that are commonly used in GW analysis: PTMCMC and dynamic nested sampling (DNS). We begin with toy models to demonstrate PARIS's ability to sample complex patterns, high-dimensional multimodal distributions, and heavy-tailed distributions. Subsequently, we investigate PARIS's effectiveness and robustness in GW analysis by testing it on two representative cases: searching for Galactic binaries (GBs) in LISA data \cite{Littenberg:2011zg,Littenberg:2020bxy} and performing parameter estimation for the GW150914 event \cite{LIGOScientific:2016vbw}.

PARIS has several hyperparameters (see Table~\ref{tab:notation} in Supplementary Information~\ref{app:notation}), of which three should be loosely tuned to the problem at hand. First, the global exploration parameter $N_\text{LHS}$ depends on the ratio between the size of the modes and the volume of the prior region, and so different estimates are used for the experiments in this section. Second, as a process is exploiting a mode locally, the number of seeds $N_\text{seed}$ is empirically set to approximately 10 times the expected mode number, but this number must also be estimated for each experiment. Finally, we use a diagonal initial covariance $\Sigma_\text{init}$ with conservatively small scales (and in most cases $\Sigma_\text{init}\propto I_p$, where $I_p$ is the $p\times p$ identity matrix), to ensure stable early-stage exploration by minimizing mode jumping. In future GW analyses, the Fisher information matrix \cite{vallisneri2008use} could provide a more tailored $\Sigma_\text{init}$.

In our experiments, we focus on the number of function evaluations, since it dominates the sampling cost in GW scenarios. For clarity, the initial LHS evaluations ($N_\text{LHS}$) are fully integrated into the reported function evaluations. For fair comparisons in toy model experiments, we run all compared samplers until they produce the same specified number of posterior samples. In contrast, for the GW cases, we allow different posterior sample numbers because PARIS generates a sample per iteration, and terminating with a small posterior sample number could lead to the insufficient sampling of complex, high-dimensional patterns. For the GB search, we run PARIS until the cluster number becomes stable. For GW150914 parameter estimation, we run PARIS until the evidence estimation becomes stable.

When estimating the evidence, PARIS uses the latest half of the samples in Eq.~\eqref{eq:evidence} to ensure a stable estimate based on well-converged samples. Evidence estimation is already a defining feature of nested sampling, and is thus provided in the particular implementation of DNS that we use (the Dynesty package \cite{Speagle_2020,sergey_koposov_2024_12537467}). In the case of PTMCMC, we employ the implementation in the Eryn package \cite{Karnesis:2023ras,michael_katz_2023_7705496,2013PASP..125..306F}, from whose tempered chains we compute the evidence with the stepping-stone estimator \cite{PhysRevD.99.084006}.

The tables below report each sampler's effective sample size (ESS), computed from its own stored output (Kish ESS $(\sum_t w_t)^2/\sum_t w_t^2$ for PARIS and nested sampling, chain length over integrated autocorrelation time for the tempered chains) and the ratio of ESS to likelihood evaluations. Under this metric, PARIS outperforms the alternatives across all three benchmark problems by factors ranging from 4 to 50.

The temperature spacing is determined by the \texttt{Eryn} package, which constructs and adapts the ladder automatically \cite{10.1093/mnras/stv2422}; we manually configure only the numbers of rungs and walkers, set empirically for each problem: five temperatures for the tempered runs (the 2D and 10D toy problems, and the GB search) and one otherwise. This setup employs fewer rungs than the package's own tutorials, and because the reported likelihood count sums evaluations across all temperatures, this choice conservatively understates the baseline sampler's computational cost rather than inflating it. A similar conservative accounting applies to nested sampling: live points are the package default of $500$ on the easiest problems and $1000$--$5000$ on the harder ones, and our stopping tolerances are looser than its built-in $0.01$. Both comparators had the cost-favourable setting. Tuning such hyperparameters is itself computationally demanding, and the requisite exploratory runs incur an overhead that is typically omitted from standard evaluation counts.

\subsection{Generic Test Problems}
\label{subsec:generic}

\subsubsection{NUS Model}

\begin{figure}[tbp]
    \centering
    \includegraphics[width=1\textwidth]{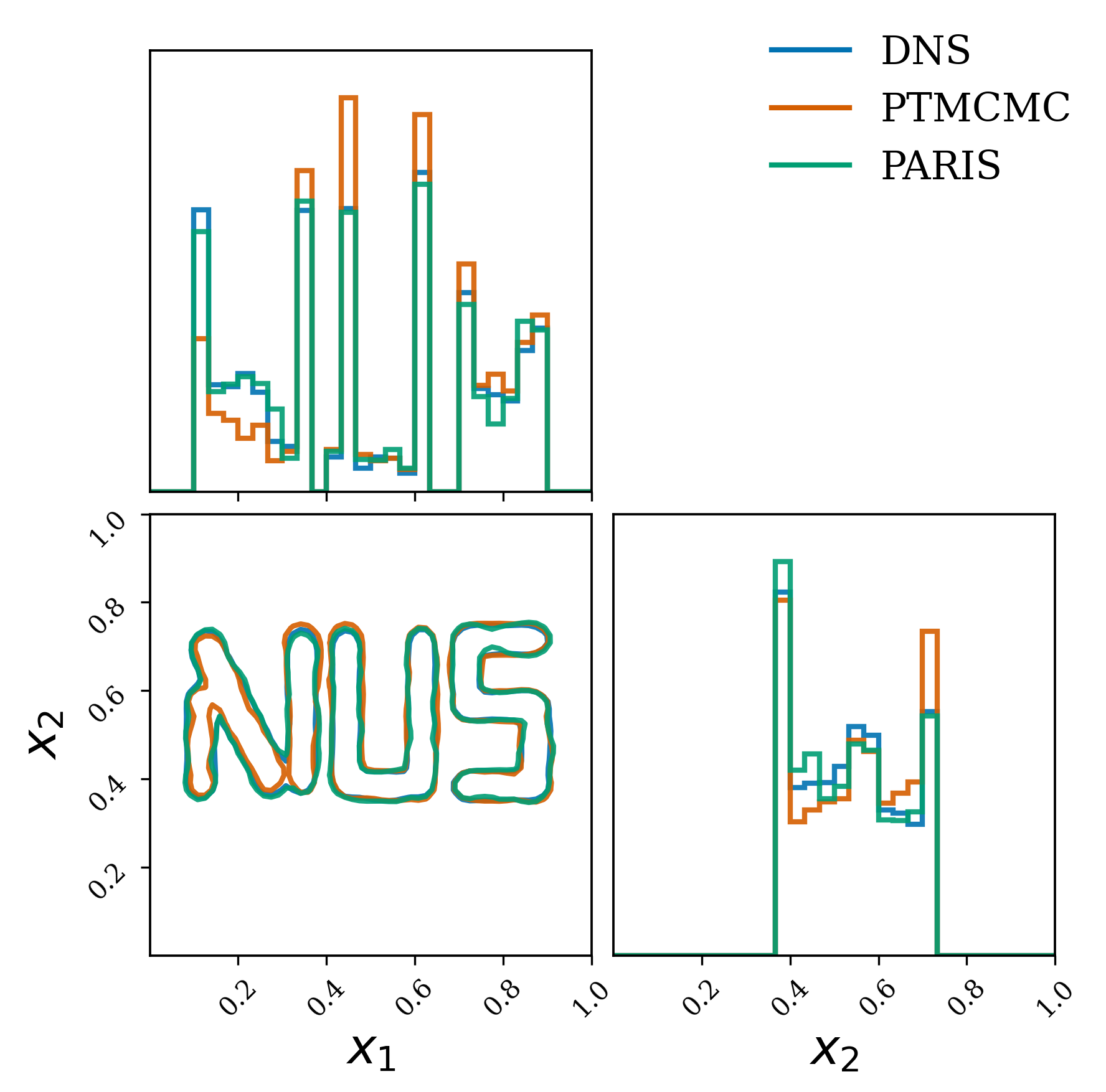}
    \caption{Reconstructed posterior distributions for the 2D ``NUS'' toy model. The target density features a complex, disjoint geometry with sharp boundaries. Green, blue and orange contours denote results from PARIS, DNS and PTMCMC, respectively. The visual comparison confirms that PARIS accurately resolves the disconnected modes and non-Gaussian features, yielding a density profile consistent with the baseline samplers.}
    \label{fig:nus_model}
\end{figure}

To evaluate the sampling performance on complex posterior landscapes, we consider a 2D toy distribution whose support traces out the letters ``NUS''---a structured yet irregular pattern composed of disconnected regions, curved boundaries, and sharp edges. This example serves as a controlled testbed for assessing the sampler's ability to resolve non-Gaussian, multi-modal distributions with minimal assumptions.

This test challenges the core assumptions of many standard methods. For instance, DNS relies on an exponential contraction of the prior volume, which becomes inefficient when the modes are not nested in a simple hierarchical structure. PTMCMC, while capable of escaping local modes through temperature ladders, often requires a large number of function evaluations to sufficiently mix across modes---especially when the posterior has narrow bridges or disconnected components.

The sampling efficiency and performance of the three methods on the 2D ``NUS'' toy model are summarized in Figure~\ref{fig:nus_model} and Table~\ref{tab:nus_model_results}. PARIS successfully reconstructs the complex geometry of the target density using significantly fewer function evaluations, with hyperparameters set to $N_\text{LHS} = 100$, $N_\text{seed} = 1$, and $\Sigma_\text{init} = 0.5 \times I_p$. In contrast, DNS requires a much higher computational budget to achieve similar resolution, configured with 500 live points and a stopping criterion of $\texttt{dlogz}_{\text{init}} = 0.05$. Similarly, the PTMCMC run, utilizing 10 walkers across 5 temperatures with a 1000-step burn-in period, also exhibits lower efficiency compared to the PARIS framework.

\begin{table}[tbp]
    \centering
    \begin{tabular}{lccccc}
    \hline
    Method & Posterior Samples & Function Evaluations & ESS & ESS/Evaluation & Log Evidence \\
    \hline
    DNS    & 10000 & 1500000 & 10100 & $6.7\times10^{-3}$ & -2.40 \\
    PTMCMC & 10000 & 87000   & 2980 & $3.4\times10^{-2}$ & -2.5 \\
    PARIS  & 10000 & 10100   & 1370 & $1.4\times10^{-1}$ & -2.45 \\
    \hline
    \end{tabular}
    \caption{Performance metrics for the 2D ``NUS'' toy model. PARIS achieves the same posterior resolution as DNS and PTMCMC, with significantly fewer function evaluations.}
    \label{tab:nus_model_results}
\end{table}

\subsubsection{10D GMM}

We further evaluate the performance of PARIS in a higher-dimensional ($p=10$) setting involving a GMM with equally weighted modes. This example is designed to test the algorithm’s scalability and robustness in higher-dimensional and strongly multi-modal spaces. The centers of the 10D GMM are generated using LHS, which ensures that the modes are well separated across all dimensions. This placement strategy creates a challenging target distribution for samplers: although some modes appear close in the one-dimensional marginals, they are in fact separated by large low-likelihood regions in the full 10D space. As a result, mode identification and exploration become non-trivial for DNS, which relies on clustering in live-point space, as well as for PTMCMC, which requires efficient mixing across modes.

\begin{figure}[tbp]
    \centering
    \includegraphics[width=\textwidth]{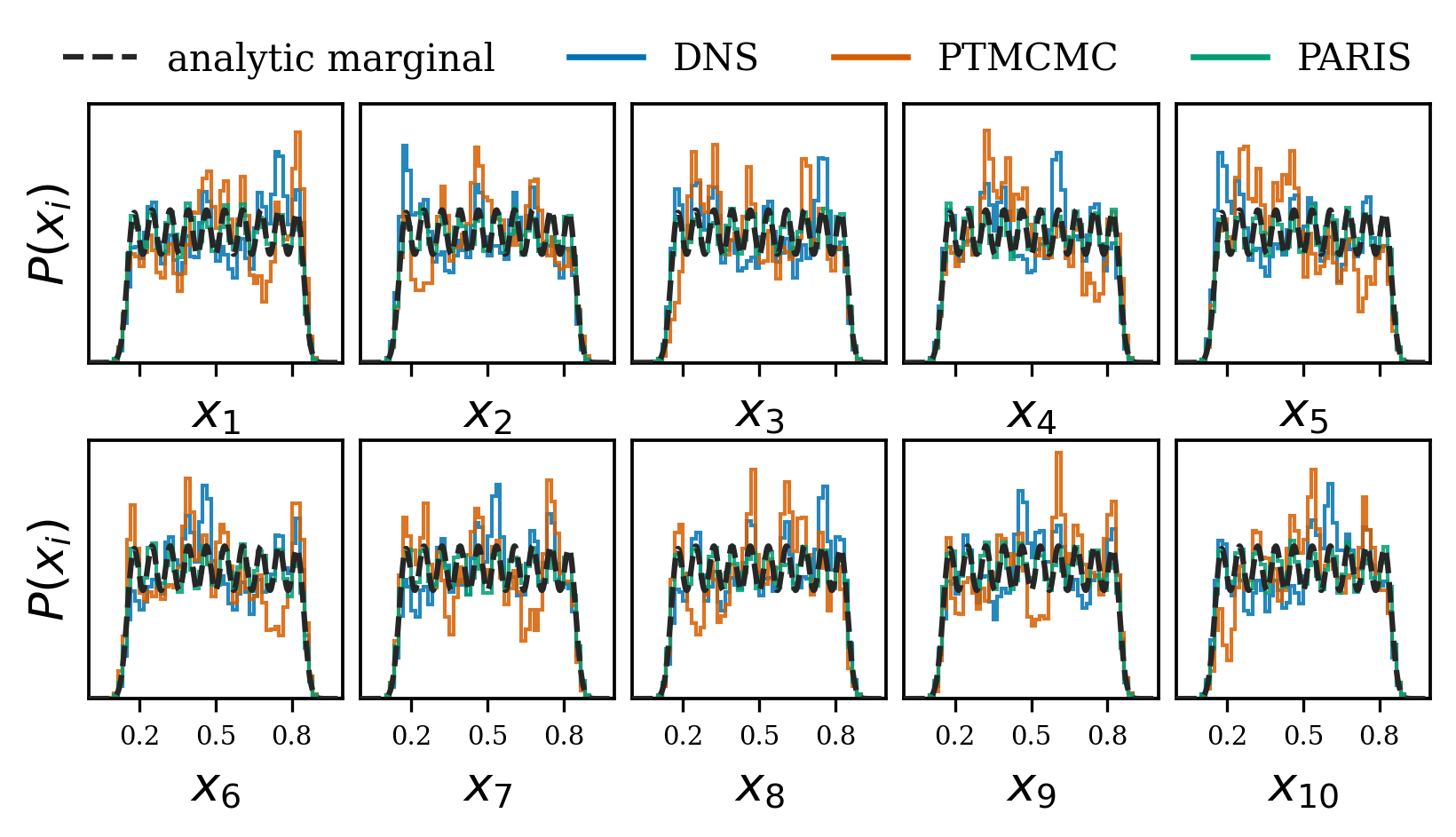}
    \caption{Comparison of 1D marginalized posterior distributions for a 10-dimensional GMM with 10 equally weighted modes, whose centers are selected via LHS to ensure maximal separation. The black dashed curve represents the analytically computed marginalized density of the target GMM, appearing uniform in 1D projections due to the mode placement. PARIS achieves consistent sampling across all modes with fewer function evaluations, closely aligning with the analytical solution. In contrast, DNS and PTMCMC use more function evaluations, but show uneven mode recovery.}
    \label{fig:10d_modes}
\end{figure}

In our tests, the PARIS hyperparameters are \( N_\text{LHS} = 10^4 \), \( N_\text{seed} = 100 \), and \( \Sigma_\text{init} = 10^{-3} I_p \). DNS is configured with 5,000 initial live points and a stopping criterion of $\texttt{dlogz}_\texttt{init} = 0.05$. PTMCMC uses 100 walkers, 5 temperatures, and 1,000 burn-in steps. From Figure~\ref{fig:10d_modes}, we observe that the modes in the target distribution are all of equal weight, and PARIS yields consistent results across them. In contrast, both DNS and PTMCMC exhibit significant biases among modes. Table~\ref{tab:10d_gaussian_modes} compares the posterior evaluation for samplers and their evidence estimation. Both PARIS and DNS give accurate evidence estimation, though DNS uses many more posterior evaluations. In Figure~\ref{fig:10d_modes}, DNS exhibits a peak height bias in the 1D marginalized likelihood distributions. Despite this local discrepancy, DNS provides accurate evidence estimates. This accuracy arises because, in high-dimensional Gaussian distributions, the majority of the probability mass is concentrated in a shell at a radius not near the center where the bias is most apparent. Consequently, DNS effectively captures the dominant contribution to the evidence integral from the outer regions of each GMM component. PARIS not only achieves accurate sampling but also does so with substantially fewer function evaluations than DNS and PTMCMC.


\begin{table}[tbp]
    \centering
    \begin{tabular}{lccccc}
    \hline
    Method & Posterior Samples & Function Evaluations & ESS & ESS/Evaluation & Log Evidence \\
    \hline
    DNS & 145423 & 8587847 & 40594 & $4.7\times10^{-3}$ & 2.30 \\
    PTMCMC & 145400 & 822352 & 7657 & $9.3\times10^{-3}$ & 1.80 \\
    PARIS & 145420 & 160050 & 35239 & $2.2\times10^{-1}$ & 2.30 \\
    \hline
    \end{tabular}
    \caption{Quantitative comparison on a 10-dimensional target distribution GMM with 10 equally weighted modes. PARIS recovers the exact log-evidence ($\ln Z = 2.30$), matching DNS but requiring approximately 50 times fewer function evaluations. PTMCMC fails to converge to the true evidence value despite a higher computational cost.}
    \label{tab:10d_gaussian_modes}
\end{table}

\subsubsection{10D Crossed Ridges: Edge Cases of the Merge Rule}\label{subsec:xshape}

To systematically evaluate the merge rule of Sec.~\ref{sec:Base Implement} under the edge cases identified earlier, we construct a 10-dimensional benchmark target that instantiates all of these challenging features simultaneously. The target is a two-component GMM with elongated, mutually intersecting modes whose evidence and relative mode weights are analytically tractable; Table~\ref{tab:xshape_cases} maps each edge case to the corresponding structural property that realizes it.

\begin{table}[tbp]
    \centering
    {
    \begin{tabular}{ll}
    \hline\hline
    Edge case & Instantiation in benchmark \\
    \hline
    Ridges & Elongated, narrow Gaussian ridges \\
    Strong correlations & Principal axes tilted at $25^\circ$ from coordinate axes \\
    Unequal-volume modes & Effective mode volume ratio of $56$ \\
    Bridging modes & Density overlap forming a continuous bridge at the intersection \\
    Peak height $\neq$ integrated mass & Mode with $2.6\times$ higher peak carries $21.6\times$ lower integrated mass \\
    \hline\hline
    \end{tabular}}
    \caption{Edge cases of the merge rule, and the properties of the 10-dimensional target that instantiate each scenario. The inversion of peak height and integrated posterior mass represents the configuration where selecting by peak height could, in principle, bias the evidence and mode weight estimates.}
    \label{tab:xshape_cases}
\end{table}

\begin{figure}[tbp]
    \centering
    \includegraphics[width=\textwidth]{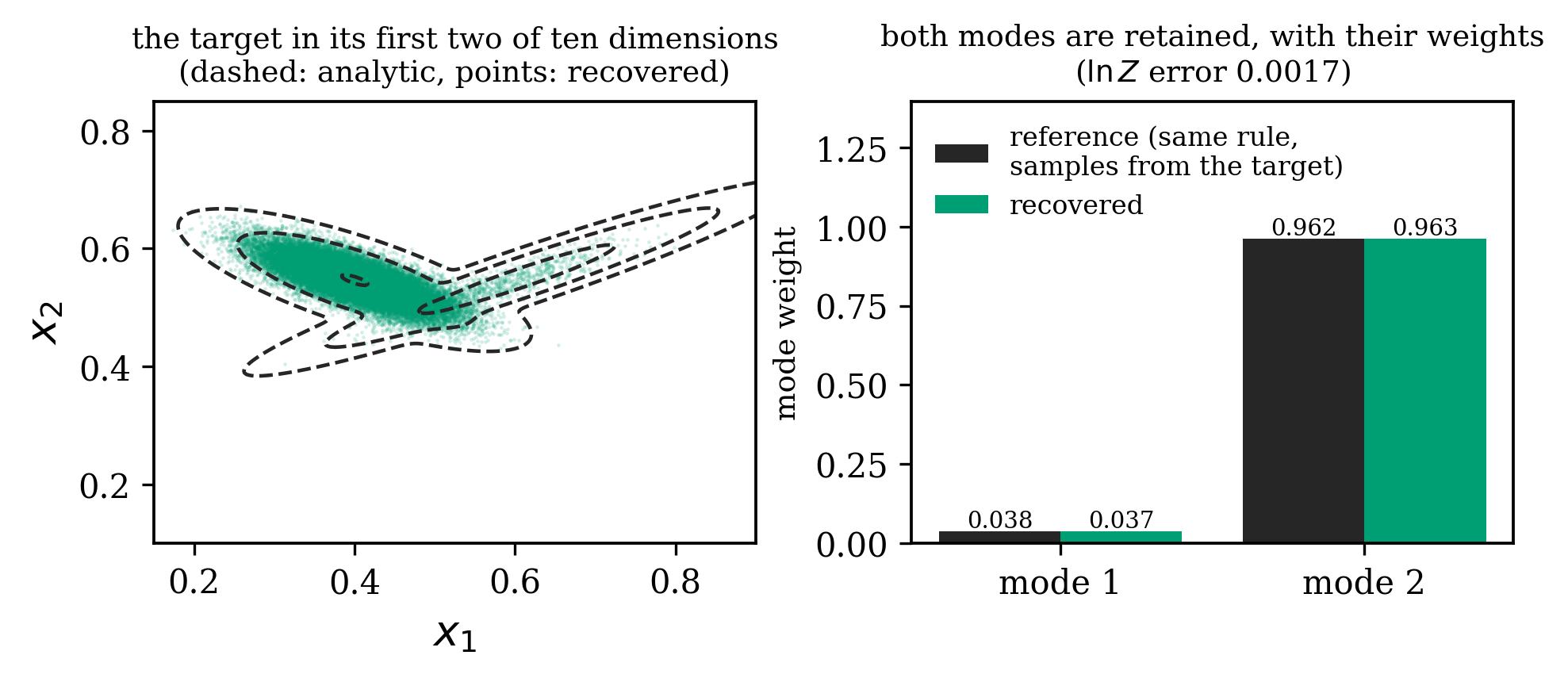}
    \caption{The target of Table~\ref{tab:xshape_cases}, and what PARIS recovers on it. \emph{Left}: the target is 10-dimensional, and its first two coordinates are shown with the other eight held at the centre of the prior; the dashed contours are the analytic density and the points are the recovered samples. \emph{Right}: the weight of each mode. The reference bar is the same mode-assignment rule applied to samples drawn directly from the target rather than the analytic weight, since the rule cannot divide the region where the two modes overlap.}
    \label{fig:xshape}
\end{figure}

We run PARIS on this target with the same settings as the other toy problems ($N_\text{LHS} = 10^4$, $N_\text{seed} = 100$, $\Sigma_\text{init} = 10^{-3} I_p$, $\alpha = 10^3$). In each of five random seeds the $100$ processes merge into one and both modes are retained. The run reported here produces $5\times10^5$ posterior samples, and the evidence is recovered to $0.002$ in $\ln Z$ using $5.1\times10^5$ function evaluations, against $4.6\times10^6$ for DNS. To estimate the relative weight of each mode from the samples of the surviving process, we compute the Mahalanobis distance from each sample to both mode centres with respect to their respective covariances, assign the sample to the nearer mode, and sum the corresponding importance weights within each partition. Applying this exact assignment rule to $4\times10^5$ independent samples drawn directly from the target yields a reference weight of $0.038$ for the subdominant mode, against its analytical integrated mass of $0.044$. The difference is the region where the two modes overlap, which no such rule can divide. PARIS gives $0.037$, and both are shown in Fig.~\ref{fig:xshape}.

\subsection{Gravitational-Wave Applications}\label{subsec:gwapps}

The remaining experiments concern the setting that motivates this work. The advancement of GW detection and modeling has led to increasingly multi-modal posteriors. Next-generation detectors (LISA \cite{LISA:2017pwj,Amaro-Seoane:2012vvq}, Tianqin \cite{TianQin:2015yph,Li:2024rnk}, Taiji \cite{Hu:2017mde}, Einstein Telescope \cite{ET:2019dnz,Punturo:2010zz}, Cosmic Explorer \cite{Evans:2021gyd,LIGOScientific:2016wof}), with their enhanced sensitivity and frequency coverage, will increase the likelihood of overlapping signals and thus require the simultaneous detection of multiple sources. Comparing a single-source template to multi-source data naturally results in a multi-modal likelihood, as different parameter regions can partially explain the data \cite{OShaughnessy:2017tak}. Additionally, complex waveforms like extreme mass-ratio inspiral signals contain numerous secondary peaks, further complicating sampling by obscuring the primary peak \cite{Chua:2022ssg,Chua:2021aah}. At the same time, GW parameter inference remains a high-dimensional problem. The dimensionality for a single source typically ranges from 10 to 20 \cite{Christensen:2022bxb}, while LISA global fitting requires simultaneously estimating thousands of overlapping sources plus noise parameters, creating joint parameter spaces with tens to hundreds of thousands of dimensions that necessitate global rather than sequential analysis approaches \cite{Cornish_2005,Littenberg_2023}. In pulsar timing array studies, the parameter space dimensionality is also inflated due to the need to simultaneously infer detailed noise models for each individual pulsar \cite{Antoniadis_2022,Xu2023SearchingFT}.

\subsubsection{Toy Overlapping GW Signal}

To evaluate the robustness of PARIS in the presence of long tails and multiple modes, we construct a toy GW inference problem inspired by the slowly evolving, low-eccentricity GB signals that LISA will observe \cite{Robson_2018}. The waveform is modeled as a sinusoidal chirp of the form:
\begin{equation}
h(t; x) = \sin(\phi + f t + \dot{f} t^2 + \ddot{f} t^3),
\end{equation}
where \( x = (\phi, f, \dot{f}, \ddot{f}) \) defines the signal parameters. Despite its simplicity, this model captures essential features of real waveforms (such as frequency evolution) while also inducing complex likelihood structures.

In this 4-dimensional example, we take the superposition of two signals as the data stream \( s(t) \), each with distinct parameters but within the same prior volume (re-scaled to the unit hypercube \([0,1]^4\)). This construction creates a controlled multi-modal posterior with two well-separated peaks. Crucially, the posterior exhibits long tails and multiple local maxima, as is the case in more realistic GW posteriors. These arise from the fact that small changes in frequency or phase can yield partially aligned waveforms, producing aliased peaks and ridges in the likelihood. These structures pose challenges to traditional samplers---the movement of PTMCMC is hindered by multiple local maxima, while DNS needs more live points to sample the tail regions correctly.

In the context of GW astronomy, the strain data $s(t)$ (representing the general data $D$ in Sec. \ref{sec:background}) collected by an interferometer is modeled as a combination of a deterministic signal $h(t; x)$ and stochastic noise $n(t)$:
\begin{equation}
s(t) = h(t; x) + n(t).
\end{equation}
Assuming the detector noise $n(t)$ is additive, stationary, and Gaussian, the probability of observing data $s$ given the source parameters $x$ is described by the Whittle likelihood \cite{allen2012findchirp}:
\begin{equation}\label{gw_like}
L(x | s) \propto \exp\left( -\frac{1}{2} \langle s - h(x) \mid s - h(x) \rangle \right),
\end{equation}
where $\langle \cdot \mid \cdot \rangle$ is a noise-weighted inner product \cite{allen2012findchirp}:
\begin{equation}\label{eq:inner}
\langle a \mid b \rangle = 4 \Re \int_{f_{\min}}^{f_{\max}} \frac{\tilde{a}(f)\tilde{b}^*(f)}{S_n(f)} df.
\end{equation}

Here, $\tilde{a}(f)$ is the Fourier transform of $a(t)$, and $S_n(f)$ is the one-sided noise power spectral density (PSD) of the detector. If we assume a flat prior $\pi(x)$ over the relevant search region, the unnormalized log-posterior $\ln P(x)$ simplifies to the quadratic form:
\begin{equation}\ln P(x) = -\frac{1}{2} \langle s - h(x) \mid s - h(x) \rangle + \text{const}.\end{equation}
This assumption is used for both the simplified toy GW model and the more realistic parameter estimation tasks presented in later sections. In this toy model, we adopt a constant PSD, setting $S_n(f) = 30$ for all frequencies, which simplifies the calculation while preserving the essential behavior of the likelihood. The simulated data stream $s(t)$ has a duration of $100\,\text{s}$, sampled at $10\,\text{Hz}$.

\begin{figure}[tbp]
    \centering
    \includegraphics[width=1\textwidth]{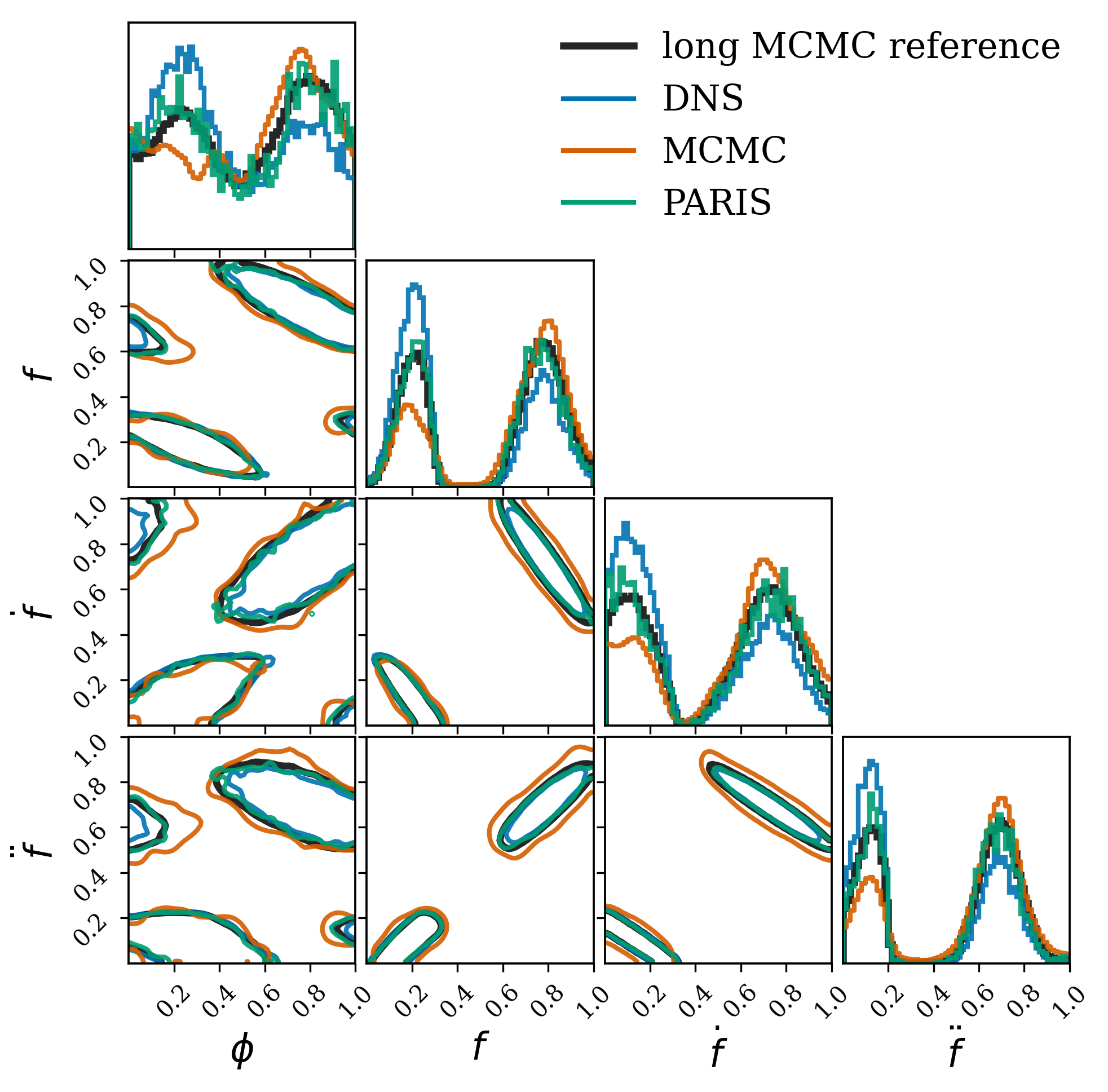}
    \caption{Corner plot comparison of the 4D toy GW posterior, parameterized by phase ($\phi$), frequency ($f$), and its derivatives ($\dot{f}, \ddot{f}$). The reference distribution (thick black) is derived from an extensive MCMC run. PARIS correctly reproduces the complex multi-modal structure and non-linear correlations, matching the reference. In contrast, both DNS and a shorter MCMC run exhibit significant sampling error.}
    \label{fig:toygw}
\end{figure}

We run PARIS with hyperparameters \( N_\text{LHS}=10^3 \), \( N_\text{seed}=100 \) and \( \Sigma_\text{init}=10^{-3} I_p \). For comparison, DNS is configured with multi-bound sampling, the \texttt{rslice} sampler, 500 live points, and a stopping criterion of $\texttt{dlogz}_\texttt{init} = 0.05$. MCMC is configured with 10 walkers, 1 temperature, and 100 burn-in steps as an intuitive configuration for this experiment. We tried several numbers of temperatures for this comparison and found that a single temperature reached comparable performance at the smallest number of function evaluations, which is why no tempered run is used as the comparator here; the cold chain already moves between the modes of this posterior, so the extra temperatures buy little for their cost. Figure~\ref{fig:toygw} and Table~\ref{tab:toygw} present the comparative results. The thick black contour in Figure~\ref{fig:toygw} represents a reference solution from a long-running MCMC run with 6.5 million function evaluations, using 10 walkers at a single temperature, as for the shorter run above. PARIS recovers the full structure of the posterior more accurately than DNS and the short MCMC run, as evidenced by the 1D and 2D marginal posteriors. It also achieves better sampling efficiency, requiring fewer function evaluations than both other methods. PARIS shows similar evidence estimation results to DNS.

\begin{table}[tbp]
    \centering
    \begin{tabular}{lccccc}
    \hline
    Method & Posterior Samples & Function Evaluations & ESS & ESS/Evaluation & Log Evidence \\
    \hline
    DNS & 15213 & 510089 & 11636 & $2.3\times10^{-2}$ & -18.40 \\
    MCMC & 15200 & 106688 & 1595 & $1.5\times10^{-2}$ & -19.30 \\
    PARIS & 15200 & 26297 & 3176 & $1.2\times10^{-1}$ & -18.49 \\
    \hline
    \end{tabular}
    \caption{Performance metrics for the 4D toy GW model. All samplers were run to generate approximately $1.5 \times 10^4$ posterior samples. PARIS yields a log-evidence estimate consistent with DNS ($\ln Z \approx -18.4$), but reduces the computational cost by a factor of $\approx 20$. In contrast, MCMC exhibits a significant evidence bias, indicating incomplete exploration of the multi-modal posterior.}
    \label{tab:toygw}
\end{table}

\subsubsection{Searching for GB Signals in LISA Data}

The Laser Interferometer Space Antenna (LISA) is a planned space-based GW observatory with arm lengths significantly longer than those of ground-based detectors, enabling high sensitivity in the millihertz frequency band \cite{LISA:2017pwj,Amaro-Seoane:2012vvq}. In this band, the dominant sources are expected to be stellar-mass binaries in the Milky Way (primarily white-dwarf binaries) which emit nearly monochromatic GWs throughout the mission duration. Due to their abundance, LISA is expected to observe millions of such GB systems \cite{LISACosmologyWorkingGroup:2022jok,LISA:2022yao}. However, only about $\sim 10^4$ of these binaries will be individually resolvable. The remaining unresolved sources will form a confusion foreground, resulting in overlapping signals in both the time and frequency domains. Disentangling tens of thousands of overlapping signals over the mission lifetime presents a substantial data analysis challenge.


Resolving individual sources generally involves two main phases: searching and parameter estimation. During the search phase, we aim to identify the maximum a posteriori (MAP) point in parameter space for each source. Parameter estimation then refines the posterior distribution once the signals have been localized in that space. In this experiment, we focus on the search task to demonstrate the performance of PARIS on optimization, as the posterior landscape contains numerous secondary modes that can trap traditional samplers in their search for the primary mode.

Existing search methods in GW data analysis include coordinate descent \cite{Strub_2022}, swarm-based algorithms \cite{Zhang:2021htc,Bouffanais:2015sya} and extensively used MCMC
\cite{Crowder:2006eu, Littenberg:2011zg, Littenberg:2020bxy, Littenberg:2023xpl}.
Recently, a versatile MCMC sampler called Eryn was introduced \cite{Karnesis:2023ras}, designed to unify many modern Bayesian inference strategies (such as trans-dimensional MCMC) into a single framework. Eryn has emerged as a significant tool for tackling complex inference tasks, such as the LISA global fit problem, making it a key point of comparison in this experiment. However, the search phase still poses distinct challenges due to the high degree of multi-modality in the problem. To address these challenges, we leverage the intrinsic adaptive mechanism of PARIS, as detailed in Sec.~\ref{subsec:Adaptive_Strategy}. This property enables the algorithm to navigate complex, multi-modal landscapes effectively, and to locate the MAP estimates as a direct byproduct of the sampling process.

A GB signal is described by eight parameters \cite{Katz_2022}, only six of which require active numerical exploration in our setup: the remaining two are eliminated analytically via the construction of the detection statistic, as detailed when the statistic is formulated below. The six sampled parameters are specified in Supplementary Information~\ref{si:config}. We choose the waveform duration to be 4 years and focus on a narrow frequency window between 3.997 and 4\,mHz, where multiple overlapping sources are present in the LDC2 catalog \cite{le_jeune_2022_7132178}, providing a sufficiently representative test case for our search algorithm. Our synthetic data set is the sum of five noiseless waveform templates:
\begin{equation}
s = h_1 + h_2 + \dots + h_5,
\end{equation}
whose parameters are selected and adapted from the LDC2 catalog in this frequency range, except that the waveform amplitudes are adjusted to yield a detection signal-to-noise ratio (SNR) of $\langle s \mid h\rangle/\sqrt{\langle h \mid h\rangle} = 40$ against the full five-source data stream. The optimal SNRs $\sqrt{\langle h \mid h\rangle}$ of the five sources in isolation are then $41.02$, $41.08$, $39.83$, $38.96$ and $39.00$; the deviations from $40$ arise directly from the mutual waveform overlaps among the five adjacent GB signals, and vanish when each template is matched solely against its own isolated waveform. This choice is a simplifying assumption to improve the performance of samplers (not just PARIS) on this illustrative example. A more realistic setup with significantly different SNRs would likely necessitate the joint analysis of all signals' parameters using a multi-source (either fixed or trans-dimensional) model; but see also the discussion in Sec.~\ref{sec:Discussion}.

We define the unnormalized log-``posterior'' for the search problem as
\begin{equation}
\log P_\mathrm{GB}(x) = \frac{\langle s \mid h(x)\rangle}{\sqrt{\langle h(x) \mid h(x)\rangle}},
\end{equation}

where the inner product $\langle\cdot|\cdot\rangle$ is now redefined as the sum of \eqref{eq:inner} over the noise-orthogonal $A$ and $E$ channels, and the PSD  is the SciRDv1 analytic model \cite{LISA_SciRD} as implemented in the LISA Analysis Tools package \cite{katz_lisatools} :
\begin{equation}
\langle a \mid b \rangle = \sum_{I \in \{A, E\}} 4 \Re \int_{f_{\min}}^{f_{\max}} \frac{\tilde{a}_I(f)\tilde{b}^*_I(f)}{S_{n,I}(f)} \, df.
\end{equation}

The $A$ and $E$ channels represent two of the three noise-orthogonal time-delay interferometry (TDI) observables for LISA. In contrast to the raw laser links, their noise components are mutually uncorrelated; consequently, the likelihood factorises across them and the inner product above reduces to the sum of the two channel contributions. The third combination, $T$, is insensitive to gravitational waves in this frequency regime. This log density is a standard GW statistic called the detection SNR; in the presence of a signal and Gaussian noise, it is normally distributed with a mean equal to the optimal SNR and unit variance \cite{allen2012findchirp}. Under this definition, the overall amplitude \(\mathcal{A}\) no longer appears in the posterior, and we additionally maximize over the phase \cite{prix2007statistic}. Consequently, the model has six free parameters; the search ranges for these are given in Supplementary Information~\ref{si:config}. All GB waveforms used here are generated with the GBGPU package \cite{michael_l_katz_2022_6500434,Cornish_2007,Robson_2018,Katz:2022izt}.

\begin{figure}[tbp]
    \centering
    \includegraphics[width=1\textwidth]{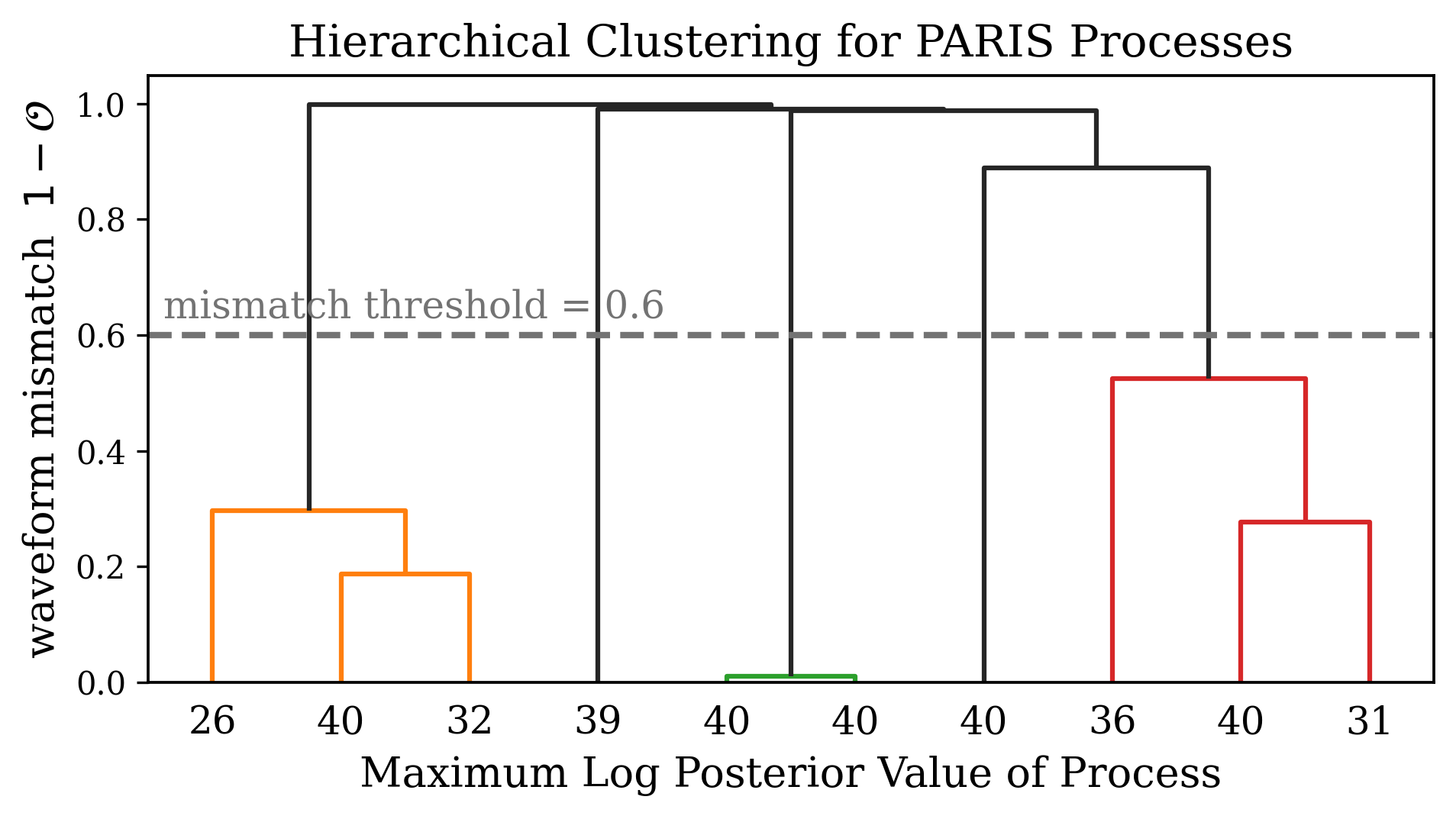}
    \caption{Dendrogram of the 10 surviving PARIS processes (reduced from 50 initial seeds) clustered by waveform mismatch ($1 - \text{Overlap}$). A broad mismatch threshold band of 0.6 (dashed line) to 0.8 resolves five distinct clusters, each corresponding to an injected signal.}
    \label{fig:gb_clustered}
\end{figure}

The PARIS hyperparameters are $N_\text{LHS}=10^4$ and $N_\text{seed}=50$; the initial covariance matrix $\Sigma_\text{init}$ is diagonal, with diagonal elements corresponding to frequency, ecliptic longitude, and ecliptic latitude set to $10^{-4}$, and all others set to $10^{-2}$. This larger scale is chosen for parameters that are expected to be weakly constrained by the data, ensuring the initial proposal sufficiently covers their broad posterior support. During the sampling process, the 50 initial processes merged into 10. The entire search yielded 100,000 posterior samples with just under 200,000 function evaluations. To identify distinct GW sources from the 10 remaining processes, we re-clustered them using the overlap between their waveforms corresponding to their MAP parameters:
\begin{equation}
    \mathcal{O}(h_1, h_2) = \frac{\langle h_1 \mid h_2 \rangle}{\sqrt{\langle h_1 \mid h_1 \rangle \langle h_2 \mid h_2 \rangle}}.
\end{equation}
Two distinct clustering mechanisms operate in this analysis. The first is the runtime merging scheme detailed in Sec.~\ref{sec:Practicle Implement}: every pair of live processes is compared via their proposal densities at the latest sample point; within each cluster, only the process with the highest peak is preserved, while the samples and weights of terminated processes are retained. The second is a post-processing step applied once to the 10 surviving processes: because several processes may occupy secondary maxima corresponding to the same physical source, we group them according to physical solution similarity, evaluated via the waveform overlap above between the best-fit waveforms across processes, rather than proposal geometry. The hierarchical clustering result is shown in Figure \ref{fig:gb_clustered}. The dendrogram reveals that for a broad range of thresholds (including the chosen threshold of 0.6), there are five clusters corresponding exactly to the five true signals.

\begin{figure}[tbp]
    \centering
    \includegraphics[width=\textwidth]{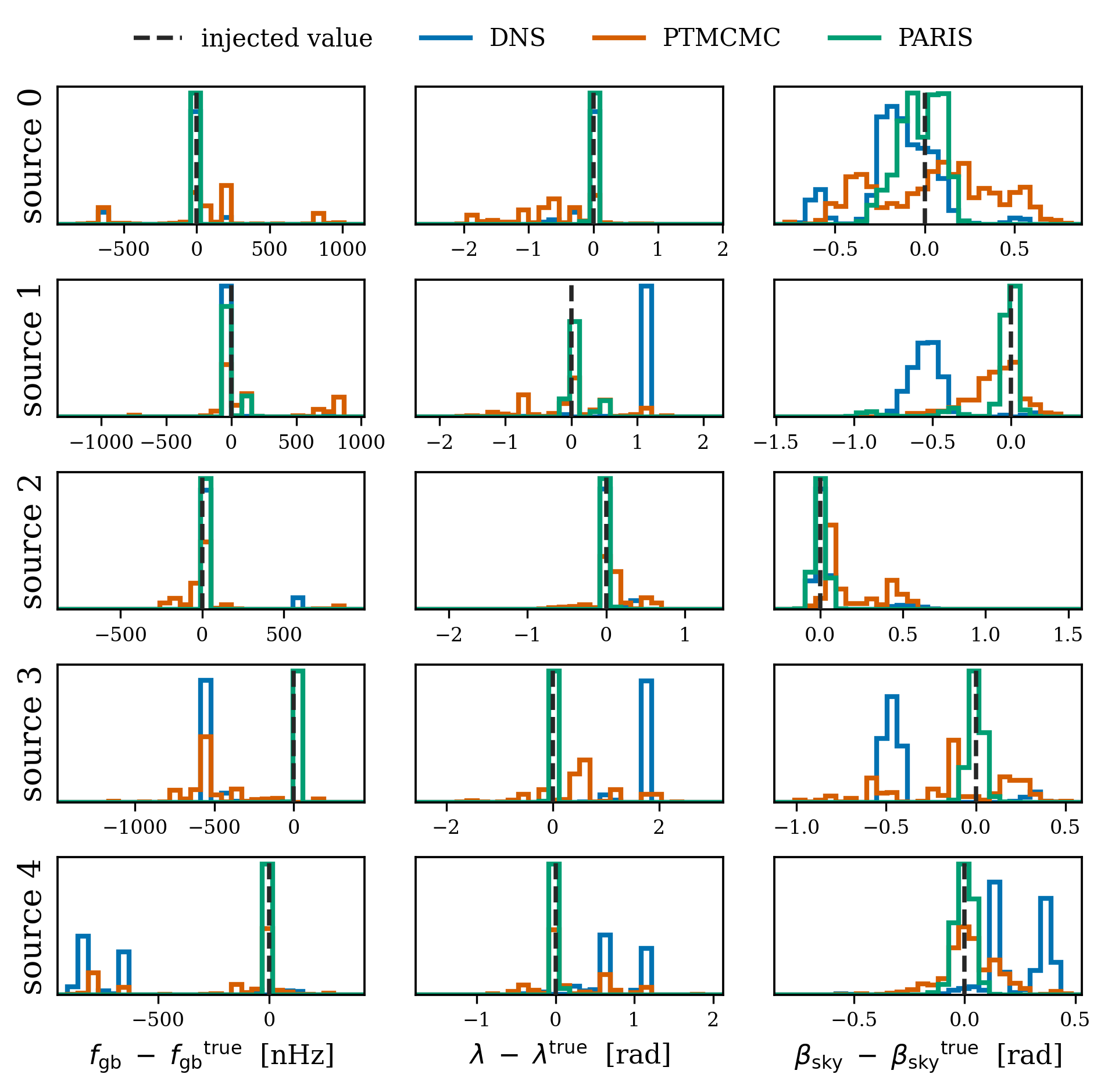}
    \caption{Posterior distributions from a 6D galactic binary signal search in simulated LISA data, with five injected GB signals. Each row corresponds to one of the five clusters identified through hierarchical clustering (see Figure \ref{fig:gb_clustered}). The results show only the well constrained parameters, with each horizontal axis describing the offset from the injected value, $x - x_\text{true}$. The green, orange and blue curves correspond to PARIS, PTMCMC and DNS, respectively. The dashed vertical line marks the injected value. PARIS successfully identifies all five injected sources, with injection parameters falling within their credible intervals. In comparison, PTMCMC misses modes 2 and 3, while DNS misses modes 1, 2 and 4.}
    \label{fig:gb_results}
\end{figure}

Using this clustering information, we analyze the parameter estimation performance for each identified source. As shown in Figure \ref{fig:gb_results}, the injection parameters are well-recovered within their respective confidence regions for all five true sources (Clusters 0-4). The posterior distributions show clear peaks at the injected parameter values, with the found MAP estimates approximately equal to the true parameters for each source (summarized in Table \ref{tab:gb_MAP}).

For comparison, we post-processed the PTMCMC and DNS results by assigning their samples to the nearest injected signal parameters. This procedure is equivalent to a single-iteration K-means clustering with centroids fixed at the true values, allowing us to explicitly count how many distinct modes each sampler successfully recovered. PTMCMC with 50 walkers (starting from the same seed points as PARIS) and 5 temperatures spent 210,000 function evaluations to obtain 64,350 posterior samples, but failed to find the signals corresponding to modes 2 and 3 in Figure \ref{fig:gb_results}. DNS, with 1,000 initial live points and a stopping criterion of $\texttt{dlogz}_\texttt{init} = 0.1$, required 1.7 million function evaluations to converge with 21,725 posterior samples, but missed modes 1, 2 and 4.

\begin{table}[tbp]
    \centering
    \begin{tabular}{lccccc}
    \hline
    & Mode 0 & Mode 1 & Mode 2 & Mode 3 & Mode 4 \\
    \hline
    DNS & 40.0 & 36.4 & 23.9 & 40.0 & 32.2 \\
    PTMCMC & 40.0 & 40.0 & 26.9 & 37.3 & 39.2 \\
    PARIS & 40.0 & 40.0 & 39.9 & 39.9 & 39.5 \\
    \hline
    \end{tabular}
    \caption{Detection SNR values corresponding to the MAP estimates for the five injected GB signals across the three tested samplers. Each signal was scaled to a detection SNR of 40 against the full data stream; its optimal SNR in isolation is given in the text. A sampler's success in capturing a mode is quantitatively demonstrated by its ability to recover detection SNRs near this value. While PARIS successfully identified all five modes, PTMCMC captured only three, and DNS captured only two.}
    \label{tab:gb_MAP}
\end{table}

\subsubsection{GW150914: The First Observed Binary Black Hole Merger}

\begin{figure*}[tbp]
    \centering
    \includegraphics[width=\textwidth]{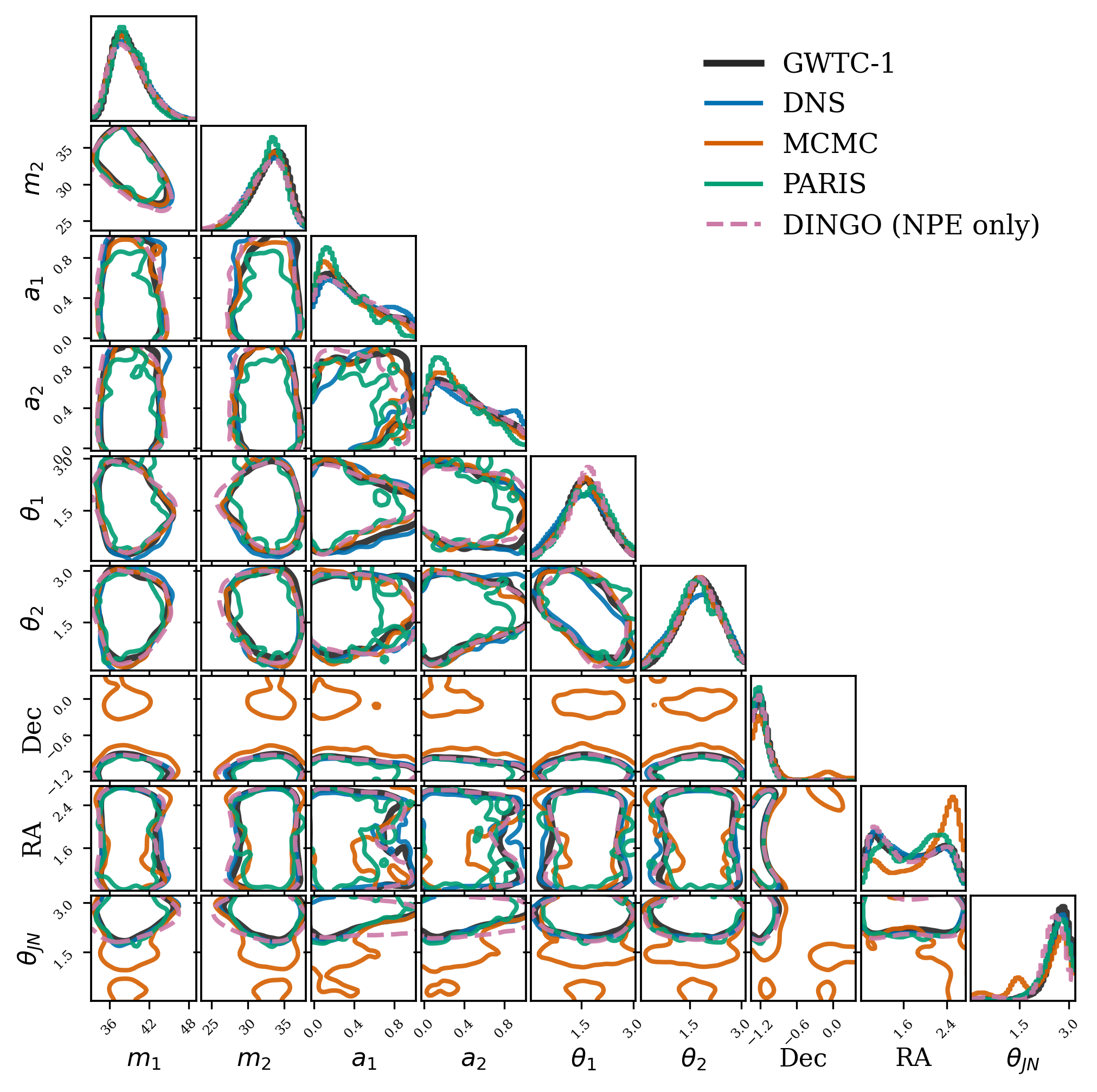}
    \caption{Corner plot for GW150914 across the nine parameters common to all five analyses: the official GWTC-1 release, our three evaluated samplers (PARIS, DNS, and MCMC), and the neural posterior estimate from Dingo-T1. Results from our three samplers are derived from the runs summarized in Table~\ref{tab:GW150914}, conditioned on the official GWTC-1 noise power spectral density, with the mass ordering convention $m_1 \ge m_2$ enforced across all distributions.}
    \label{fig:GW150914}
\end{figure*}

GW150914 was the historic first direct detection of GWs by the LIGO detectors~\cite{LIGOScientific:2016aoc}, originating from the inspiral and coalescence of two stellar-mass black holes \cite{LIGOScientific:2016vbw}. Owing to its high SNR and relatively simple structure as a quasi-circular binary black hole merger, GW150914 provides a classical example for assessing the performance of sampling algorithms in GW parameter estimation \cite{Wong:2023lgb,Green:2020dnx,Dax:2021tsq}. Despite being a single-mode distribution in the full parameter space, the posterior exhibits nontrivial internal structure and non-Gaussianity in certain parameters such as declination ($\text{dec}$), right ascension ($\text{ra}$), and inclination angle ($\theta_{\text{jn}}$).

In analyzing GW150914-like events, waveforms typically incorporate the inspiral, merger, and ringdown phases of compact binary coalescences. In our study, waveform generation is performed using the \texttt{bilby} package~\cite{bilby_paper}, which provides interfaces to various phenomenological and numerical-relativity-based waveforms. For computational efficiency, we employ an aligned-spin waveform model IMRPhenomPv2~\cite{Khan:2018fmp} which captures the essential physics of black hole mergers and is designed for robust use in data analysis.

In our parameter-estimation study of GW150914, we adopt the Hanford (H1) and Livingston (L1) strain data within a frequency band of $[20, 1024]$ Hz. The noise power spectral density (PSD) $S_n(f)$ is estimated using the Welch method with a median average and a 0.4s roll-off Tukey window. Following the framework in Sec.~\ref{sec:background}, and assuming broad uniform priors in the sampling space, the unnormalized log-posterior is equivalent to the log-likelihood defined in Eq.~\eqref{gw_like} :
\begin{equation}
\ln P(x) = \ln \mathcal{L}(x) + \text{const}.
\end{equation}
To improve sampling efficiency, we analytically marginalize over the coalescence time, phase, and luminosity distance. The resulting 13-dimensional parameter vector $x$ comprises the component masses, the spin magnitudes and orientations, the sky coordinates, the inclination and polarization angles, and a coalescence-time jitter; the complete parameter specification is provided in Supplementary Information~\ref{si:config}.

We then applied the PARIS sampler, configured with:
\[
    N_\text{LHS} = 10^4, \quad
    N_\text{seed} = 1, \quad
    \Sigma_\text{init} = 10^{-4}\,I_p.
\]
In this analysis, we initialize PARIS with a single seed at the outset, as the target posterior distribution of GW150914 exhibits a single-mode structure. PARIS completed the parameter estimation in 110 thousand function evaluations. In contrast, MCMC---configured with 10 walkers and a single temperature (since the distribution is effectively single-mode)---required 3 million function evaluations with greater sampling error (see Figure~\ref{fig:GW150914}). The number of temperatures was again chosen by trial, and a single temperature gave comparable performance at the smallest number of function evaluations, which is why no tempered run is used as the comparator here; with a posterior this close to single-mode, further temperatures are redundant and only add evaluations. DNS, initialized with 2000 live points and using a stopping criterion of $\texttt{dlogz}_\texttt{init} = 0.1$, also totaled 3 million function evaluations. Table~\ref{tab:GW150914} compares the number of posterior samples, function evaluations and log evidence for each method, with PARIS having a similar evidence estimate to DNS. The secondary mode near $\text{dec}\simeq0$ in the MCMC contours lies on nearly the same Hanford--Livingston arrival-time annulus as the main mode, and corresponds to the shorter section of that arc near the equator which is also present, with much lower probability, in the published sky map \cite{GW150914localization}. A single-temperature chain started from widely spread walkers vacates it only slowly; extending the burn-in would suppress it, at the price of discarded evaluations that would still count towards the cost reported in Table~\ref{tab:GW150914}.

\begin{table}[tbp]
    \centering
    \begin{tabular}{lccc}
    \hline
    Method & Posterior Samples & Function Evaluations & Log Evidence \\
    \hline
    DNS & 40,567 & 3,109,912 & -32108.26 \\
    MCMC & 30,000 & 3,202,895 & - \\
    PARIS & 100,000 & 110,000 & -32108.43 \\
    \hline
    \end{tabular}
    \caption{Efficiency benchmarks for the 13D GW150914 parameter estimation problem. PARIS produces a larger posterior sample set ($10^5$) while reducing the computational cost by a factor of $\sim 30$ compared to DNS and MCMC. The excellent agreement in log-evidence estimates between PARIS and DNS ($-32108.43$ vs. $-32108.26$) validates the method's accuracy in capturing the complex, correlated posterior.}
    \label{tab:GW150914}
\end{table}

As seen in Figure~\ref{fig:GW150914} and Table~\ref{tab:GW150914}, PARIS and DNS yield largely consistent posterior structures and matching log-evidence estimates. Since these are two independent sampling frameworks, their agreement indicates a converged and reliable estimation of the true posterior. With this baseline established, we observe that MCMC overestimates the probability density in low-posterior regions, as evidenced by its wider and more uneven contours (particularly in sky location and inclination). While DNS accurately reproduces the posterior, it requires millions of function evaluations to achieve such fidelity. In contrast, PARIS effectively captures the landscape with significantly fewer evaluations, but at the cost of more uneven (under-resolved) sampling in some 1D marginals; nevertheless, these do not have a big impact on the evidence estimation. Figure~\ref{fig:GW150914} also compares our results against the publicly released GWTC-1 IMRPhenomPv2 posterior.

It is instructive to contextualize SBI relative to the present work. The primary strength of SBI lies in amortization: a substantial upfront investment in simulation and network training enables near-instantaneous posterior inference for subsequent observations. In gravitational-wave astronomy, this paradigm is most prominently realized through neural posterior estimation, which now accurately reproduces sampling-based posteriors for real events \cite{Green:2020dnx,Dax:2021tsq,Kofler:2025dingo}. Conversely, its limitations are reciprocal: trained networks remain constrained to their specific training priors and model assumptions, and for several source classes motivating our work, generating an exhaustive and representative training set is itself computationally intractable. While PARIS incurs its computational cost on a per-observation basis, it requires no prior training data. Furthermore, integrating PARIS with neural proposals derived from SBI represents a promising hybrid direction for future exploration.

The figure also includes a modern simulation-based inference benchmark for this event: the neural posterior estimate from \texttt{Dingo-T1} \cite{Kofler:2025dingo}, the publicly released model of the \texttt{Dingo} framework \cite{Dax:2021tsq,Dax:2022pxd}, which couples neural posterior estimation with importance sampling and is therefore especially close in spirit to our work. At the time of writing, the released model does not incorporate \texttt{Dingo}'s importance-sampling refinement, and the network is trained on \texttt{IMRPhenomXPHM} waveforms rather than the \texttt{IMRPhenomPv2} model used here. We evaluate only the posterior distributions rather than runtime, as a per-event cost comparison against an amortized method reflects the amortization paradigm rather than the intrinsic efficiency of the sampler.

\section{Analysis and Discussion}
\label{sec:Discussion}
In a series of experiments with a range of dimensionality and posterior morphology, PARIS consistently delivers accurate parameter estimation and evidence estimation, while using significantly fewer function evaluations than standard implementations of DNS and PTMCMC. This efficiency highlights its potential as a computationally effective tool for Bayesian inference, particularly in GW analysis where function evaluations are costly. Each experiment in Sec. \ref{sec:application} showcases a unique strength of PARIS: the NUS model demonstrates its ability to resolve complex, disconnected patterns; the 10D GMM illustrates balanced mode exploitation in higher-dimensional spaces; the toy GW example emphasizes its effectiveness with long-tailed, multi-modal posteriors; the GB signal search underscores its utility for the exploration and optimization of large, complex landscapes; and the GW150914 example validates its precision in estimating a noisy and non-Gaussian single-mode posterior. These diverse challenges collectively affirm PARIS's versatility and robustness across varied inference tasks.

By using previous samples as Gaussian mixture centers weighted by their importance, PARIS focuses sampling on high-posterior regions while exploring under-sampled areas, balancing exploitation and exploration effectively. The parallel processes, initialized via LHS and refined through merging, ensure thorough coverage and reduce redundancy, while the reweighting scheme provides accurate evidence estimates. The adaptive strategy underpinning PARIS is also notably insensitive to initial conditions and hyperparameters---a feature evident across all experiments. We employed a simple diagonal initial covariance matrix \(\Sigma_{\text{init}}\) with roughly scaled diagonal elements and a fixed sliding window size \(\alpha = 1000\), yet PARIS consistently adapted to the target distribution without requiring fine-tuning.

However, PARIS has several potential limitations that can arise due to improper hyperparameter settings. 1) The presence of extremely localized modes (relative to the prior) may require more initial LHS samples or more initial processes for them to be detected reliably. 2) An excessively large $\Sigma_{\text{init}}$ may cause processes to drift erratically away from high-probability regions, whereas a $\gamma$ value that is too small can lead to numerical singularities (i.e., ill-conditioned covariance matrices) during early iterations. 3) When a single process attempts to bridge multiple adjacent modes, an insufficient sliding window size $\alpha$ may fail to construct a proposal capable of enveloping the multi-peaked landscape.

To improve the PARIS algorithm, several future directions are promising. First, incorporating location-dependent information---where the proposal geometry adapts as a function of its position (e.g., via analytical kernels or neural networks)---would enable a single process to more effectively bridge and envelope multiple adjacent modes in complex landscapes \cite{Dax:2022pxd}. Second, to improve proposal efficiency, past samples could be used to train predictive models like Gaussian processes \cite{Chua:2019wgs,Moore:2015sza,Liu:2023oxw,ElGammal:2025dkz} to identify and reject low-density regions. Last, the extension of PARIS to deal with significantly varying SNRs in the GB search can be explored by switching to a traditional trans-dimensional approach \cite{Littenberg:2011zg,Littenberg:2020bxy}, but another possibility could be to introduce modifications to the base algorithm that prioritize the local exploitation of individual processes, such as reverse annealing or an overlap-based merging criterion. Finally, as noted in Sec.~\ref{sec:Base Implement}, a more sophisticated initial design would start the processes closer to the modes and so benefit both the optimisation and the sampling that follow seeding.


\section*{Code Availability}
The source code for the PARIS algorithm, along with instructions and examples used to generate the results presented in this work, are publicly available on GitHub at \url{https://github.com/mx-Liu123/parismc}.

\section*{Data Availability Statement}
The data that support the findings of this study are openly available: the GW150914 strain data from the Gravitational Wave Open Science Center at \url{https://gwosc.org}, the GWTC-1 posterior samples at \url{https://dcc.ligo.org/LIGO-P1800370/public}, and the Galactic-binary source parameters from the LISA Data Challenge catalogue \cite{le_jeune_2022_7132178}. The toy models are defined in full in the text. The PARIS implementation is publicly available at \url{https://github.com/mx-Liu123/parismc}.

\begin{acknowledgments}
ML would like to express his sincere gratitude to Kate Lee, Nandini Sahu, Joshua Speagle, and Peter Pang for their invaluable insights and constructive feedback on this work. He also thanks several colleagues at NUS for their support and stimulating discussions throughout the development of the PARIS framework, especially Soichiro Isoyama, Josh Mathews, Shubham Kejriwal, Davendra Hassan, and Rishav Agrawal. ML is supported by the NUS Research Scholarship for his PhD.

\end{acknowledgments}

\newpage

\appendix
\renewcommand{\appendixname}{Supplementary Information}

\begin{table*}[!ht]
\renewcommand{\arraystretch}{1.5}
\begin{tabularx}{\textwidth}{|c|X|}
\hline
\textbf{Symbol} & \textbf{Description} \\
\hline
$p$ & Dimensionality of the samples \\
$P(x)$ & Unnormalized target density \\
$T, t$ & The total number of iterations and the current iteration index, respectively. \\
$Z, \hat{Z}$ & The true Bayesian evidence and its empirical estimate, respectively. (Subscript $t$, e.g., $\hat{Z}_t$ or $Z_t$, denotes the estimate or cumulative weight sum at iteration $t$) \\
$x_{t}^{(j)}$ & The sample at $t$-th iteration of the $j$-th process \\
$w_{t}^{(j)}$ & Importance weight of sample $x_{t}^{(j)}$\\
$q_{t}^{(j)}(x)$ & Proposal distribution at iteration $t$ in $j$-th process \\
$\Sigma_{t}^{(j)}$ & Covariance matrix of weighted sample set evaluated at iteration $t$ in the $j$-th process. In the practical version, updated every $\gamma$ iterations and denoted as $\Sigma_{\left\lceil t/\gamma \right\rceil}^{(j)}$ \\
$y_{t}^{(j)}$ & Proposal center for generating $x_{t}^{(j)}$ \\

$q_{\text{total}}^{(j)}(x)$ & The cumulative proposal distribution for the $j$-th process, used as the importance weight denominator. \\

$\mathcal{J}$ & The set of currently active sampling processes during the dynamic merging and splitting phases. \\

$\widetilde{\mathcal{N}}$ & Modified Gaussian distribution with a rescaled peak value, introduced to prevent density inflation in high-dimensional spaces. \\

$d_{t}^{(j)}$ & Denominator of the importance weight for $x_t^{(j)}$ at iteration $t$ \\

$d_{t}^{(j' \to j)}$ & Cross-density contribution from external process $j'$ evaluated at the current sample position of process $j$. \\


\hline
\multicolumn{2}{c}{\textbf{Method hyperparameters}} \\
\hline
$N_{\text{LHS}}$ & Number of initial Latin hypercube samples covering the prior region \\
$N_{\text{seed}}$ & Number of processes in initialization \\
$\Sigma_\text{init}$ & The scale of initial proposal \\
$\gamma$ & Iteration count to update the proposal covariance, fixed as 100 for all experiments in this work \\
$\alpha$ & Truncation parameter, considering only the latest $\alpha$ iterations' samples, fixed as 1000 for all experiments in this work \\
\hline
\end{tabularx}
\caption{Notation used throughout the paper.}
\label{tab:notation}
\end{table*}


\section{Notation}
\label{app:notation}
Table~\ref{tab:notation} summarizes the notation used throughout the paper for convenience of reference.

\section{Consistency of the PARIS Estimator}
\label{app:convergence}
The base PARIS estimator for some test function $\tau$ at iteration $T$ is defined as
\begin{equation}
\hat{\tau}_T = \frac{1}{T} \sum_{i=1}^{T} w_T(x_i) \tau(x_i),
\end{equation}
where all historical samples are used and reweighted according to the entire sequence of proposal distributions up to iteration $T$. This is an example of deterministic-mixture reweighting in AMIS, which was popularized by \cite{cornuet2011adaptivemultipleimportancesampling} and formally established to have the lowest variance lower bound in \cite{Elvira_2019}. However, while AMIS algorithms (such as PARIS) yield estimators with asymptotic variance reduction, proving their strong consistency remains difficult. The most robust attempt strictly requires the per-iteration sample size $N_t$ to grow rapidly (i.e., $\sum 1/N_t < \infty$) \cite{marin2017consistency}. This approach is not taken in PARIS (where $N_t=1$; see remark in Algorithm 1), as it is computationally unfeasible for expensive target densities such as astrophysical posteriors.

Conversely, a parallel branch of literature has successfully proven asymptotic consistency for AIS (not AMIS) with a fixed sample size per iteration, without relying on the variance reduction of deterministic-mixture reweighting. By evaluating a sample solely against the single proposal that generated it (thus preserving the martingale property), \cite{portier2018asymptotic} and \cite{delyon2021safe} established uniform convergence rates for fixed sample sizes. Optimization-based perspectives further proved fixed-sample convergence by formulating parametric AIS as a stochastic gradient descent problem \cite{akyildiz2021convergence, akyildiz2024globalconvergenceoptimizedadaptive}. However, these fixed-sample AIS results expose the difficulty of establishing asymptotic consistency for target densities with non-compact support. In such cases, if the proposal's tails decay faster than the target's, the variance of the importance weights explodes to infinity. To counteract this, algorithms are often forced to incorporate artificial, heavy-tailed ``defensive'' mixtures, drastically complicating the methodology \cite{delyon2021safe}.

In PARIS, the support of the target density is explicitly required to be compact, as it is mapped onto a unit hypercube $[0,1]^p$. This design constraint (along with the Gaussian-mixture proposals) ensures that the importance weights remain bounded (i.e., $\sup (P(x)/q_{\text{total}}(x)) < \infty$), and effectively removes the risk of variance explosion in a fixed-sample approach. As also noted in Appendix B of \cite{delyon2021safe}, operating within a compact support drastically relaxes convergence requirements. Taken in combination, all of these results indicate that deterministic-mixture AMIS with a fixed sample size per iteration but a compact support will generally achieve strong consistency, although we are unable to prove it for PARIS at this point.

\section{Covariance Estimation}
\label{app:Cov}

Standard sample covariance estimates can be biased near boundaries. To address this, we model the samples as being drawn from a multivariate Gaussian distribution truncated to the prior region \(\mathcal{X}\). We propose a covariance estimation scheme that minimizes the negative log-likelihood of this truncated distribution \cite{daskalakis2018efficient}.

\subsection{Truncated Gaussian Formulation}

Let the normalization constant (partition function) over the prior region \(\mathcal{X}\) be defined as:
\begin{equation}
C(\mu, \Sigma) = \int_{\mathcal{X}} \exp \left( -\frac{1}{2} (z - \mu)^\top \Sigma^{-1} (z - \mu) \right) d z.
\end{equation}
Given a set of weighted samples \(\{ (x_{t'}, w_{t'}) \}_{t'=1}^{t}\), the negative log-likelihood function \(\mathcal{L}\) for the truncated Gaussian is:
\begin{equation}
\begin{split}
\mathcal{L}(\mu, \Sigma) =\; & \frac{1}{2} \sum_{t'=1}^{t} w_{t'} (x_{t'} - \mu)^\top \Sigma^{-1} (x_{t'} - \mu) + \left( \sum_{t'=1}^{t} w_{t'} \right) \log C(\mu, \Sigma).
\end{split}
\end{equation}
Minimizing \(\mathcal{L}\) with respect to \(\Sigma\) allows us to recover the underlying covariance structure. The integral \(C(\mu, \Sigma)\) is approximated as:
\begin{equation}
C(\mu, \Sigma) \approx r \cdot (2\pi)^{p/2} \det(\Sigma)^{1/2},
\end{equation}
where \(r\) is the volume fraction of the Gaussian distribution falling within \(\mathcal{X}\). In this work, \(r\) is estimated via Monte Carlo simulation using \(N_{\text{MC}} = 10^5\) samples drawn from the untruncated \(\mathcal{N}(\mu, \Sigma)\). Note that in the practical implementation (Algorithm~\ref{algo:practical_implement}), the recorded trial count \(N_t^{(j)}\) serves as a trigger for this expensive correction, as it provides a rough historical estimate of leakage.

\subsection{Estimation Procedure}

The optimization determines a scaling vector \(\beta\) to define a rescaled covariance \(\widetilde{\Sigma}^{-1} = \mathbf{B} \Sigma_{\text{raw}}^{-1} \mathbf{B}\), where \(\mathbf{B} = \text{diag}(\beta)\). The procedure is summarized in Algorithm~\ref{algo:cov_update}.

{\footnotesize \singlespacing
\begin{algo}\label{algo:cov_update}{Covariance Estimation}

\vss \noindent \textbf{Input:}
Weighted samples \(\{ (x_{t'}, w_{t'}) \}_{t'=1}^{t}\), Recorded trial counts \(\{ N_{t'} \}_{t'=1}^{t}\), Prior region \(\mathcal{X}\), Previous mean \(\mu\).

\vss \noindent \textbf{1) Initial Estimate}
\begin{itemize}
    \item Compute raw weighted covariance \(\Sigma_{\text{raw}}\) using Eq.~\eqref{eq:cov}.
\end{itemize}

\vss \noindent \textbf{2) Boundary Check \& Branching}
\begin{itemize}
    \item Compute average trials per sample: \(\bar{N} = \frac{1}{t} \sum_{t'=1}^t N_{t'}\).
    \item Estimate fraction outside \(\mathcal{X}\): \(f_{\text{out}} \approx 1 - 1/\bar{N}\).
    \item \textbf{If} \(f_{\text{out}} \le 0.1\):
    \[
    \widetilde{\Sigma} \gets \Sigma_{\text{raw}} \quad \text{(Truncation effect negligible)}\,.
    \]
    \item \textbf{Else} (\(f_{\text{out}} > 0.1\)): Proceed to Step 3.
\end{itemize}

\vss \noindent \textbf{3) Optimization (Truncation Correction)}
\begin{itemize}
    \item Estimate precision ratio \(r\) using \(N_{\text{MC}} = 10^5\) Monte Carlo samples from \(\mathcal{N}(\mu, \Sigma_{\text{raw}})\).
    \item Initialize scaling vector \(\beta \gets \mathbf{1}\) and define \(\mathbf{B}=\text{diag}(\beta)\).
    \item Construct objective covariance: \(\widetilde{\Sigma}(\beta) = (\mathbf{B} \Sigma_{\text{raw}}^{-1} \mathbf{B})^{-1}\).
    \item Solve for optimal scaling:
    \[
    \beta^* = \arg\min_{\beta} \mathcal{L}(\mu, \widetilde{\Sigma}(\beta))\,.
    \]
    \item Update: \(\widetilde{\Sigma} \gets \widetilde{\Sigma}(\beta^*)\).
\end{itemize}

\vss \noindent \textbf{4) Stabilization}
\begin{itemize}
\item Apply Oracle Approximating Shrinkage (OAS): \\
    To address ill-conditioning, particularly with small sample sizes, we apply OAS to $\widetilde{\Sigma}$, producing the OAS estimate $\Sigma_{\text{OAS}}$, which minimizes the mean-squared error in covariance estimation \cite{Chen_2010}:
    \[
    \Sigma_{\text{final}} \gets \Sigma_{\text{OAS}}\,.
    \]
\end{itemize}

\vss \noindent \textbf{Output:} Estimated covariance \(\Sigma_{\text{final}}\).

\end{algo}
}

\section{Configuration details for the astrophysical applications}
\label{si:config}

\subsection{Galactic-binary search: parametrization and priors}

A GB signal can be described by eight parameters \cite{Katz_2022}:
\begin{equation}
\bigl\{ \mathcal{A},\, f_\mathrm{gb} \,[\mathrm{mHz}],\, \dot{f}_\mathrm{gb}\,[\mathrm{Hz/s}],\, \phi_0,\, \cos\iota,\, \psi,\, \lambda,\, \beta_{\text{sky}} \bigr\},
\end{equation}
where $\mathcal{A}$ is the characteristic amplitude, and the frequency evolution is modeled via a Taylor expansion as $f_{\text{gw}}(t) = f_{\text{gb}} + \dot{f}_{\text{gb}}t$, where $f_{\text{gb}}$ and $\dot{f}_{\text{gb}}$ represent the GW frequency and its first derivative at the reference time $t=0$, respectively. The initial phase is denoted by $\phi_0$. The spatial location and orientation of the binary relative to the ecliptic coordinate system are encoded by the ecliptic longitude $\lambda$ and latitude $\beta_{\text{sky}}$, the inclination angle $\iota$, and the polarization angle $\psi$ \cite{Robson_2018}.

\begin{table}[htbp]
    \centering
    \begin{tabular}{l r}
    \hline
    \hline
    Parameter & Prior \\
    \hline
    \hline
         &  \\
    Frequency $f_\mathrm{gb}$ & $\mathcal{U}(0.003997, 0.004) \text{Hz}$ \\
    Log10 Frequency dot $\dot{f}_\mathrm{gb}$ & $\mathcal{U}(-19, -15)$ \\
    Inclination in Radian $\iota$ & $\mathcal{U}(0, 1.8)$ \\
    Polarization angle in Radian $\psi$ & $\mathcal{U}(0,2 \pi)$ \\
    Ecliptic longitude in Radian $\lambda$ & $\mathcal{U}(0,2 \pi)$ \\
    Ecliptic latitude in Radian $\beta_\mathrm{sky}$ & $\mathcal{U}(-0.7, 1)$ \\
     & \\
     \hline
     \hline
    \end{tabular}
    \caption{Summary of the uniform prior distributions adopted for the GB signal search. The parameter space encompasses the signal frequency ($f_{\text{gb}}$) and its derivative ($\dot{f}_{\text{gb}}$), and the source geometry ($\iota, \psi, \lambda, \beta_{\text{sky}}$). The bounds are selected to cover realistic source properties within the target frequency band.}
    \label{tab:parameters}
\end{table}

\subsection{GW150914: sampled parameters}

After the analytic marginalization over coalescence time, phase, and luminosity distance described in the main text, the 13 sampled parameters are:
\begin{itemize}
    \item Intrinsic parameters: $m_1, m_2$ (primary and secondary masses); $a_1, a_2$ (spin magnitudes); $\theta_1, \theta_2$ (spin tilt angles); $\phi_{12}$ (azimuthal angle between spin vectors); and $\phi_{\text{jl}}$ (angle between total and orbital angular momentum).
    \item Extrinsic parameters: $\mathrm{RA}, \mathrm{Dec}$ (right ascension and declination); $\theta_{JN}$ (inclination angle); $\psi$ (polarization angle); and $t_\mathrm{jitter}$ (coalescence time uncertainty).
\end{itemize}

\bibliography{apssamp}

\end{document}